# Data-driven turbulence modeling in separated flows considering physical mechanism analysis

Chongyang Yan [a], Haoran Li [a], Yufei Zhang [a,†], Haixin Chen [a]

[a] *School of Aerospace Engineering, Tsinghua University, Beijing 100084, China*

---

**Abstract:**

Accurate simulation of turbulent flow with separation is an important but challenging problem. In this paper, a data-driven Reynolds-averaged turbulence modeling approach, field inversion and machine learning is implemented to modify the Spalart–Allmaras model separately on three cases, namely, the S809 airfoil, a periodic hill and the GLC305 airfoil with ice shape 944. Field inversion based on a discrete adjoint method is used to quantify the model-form uncertainty with limited experimental data. An artificial neural network is trained to predict the model corrections with local flow features to extract generalized modeling knowledge.

† Associate professor, corresponding author; Email: zhangyufei@tsinghua.edu.cn

Physical knowledge of the nonequilibrium turbulence in the separating shear layer is considered when setting the prior model uncertainty. The results show that the model corrections from the field inversion demonstrate strong consistency with the underlying physical mechanism of nonequilibrium turbulence. The quantity of interest from the observation data can be reproduced with relatively high accuracy by the augmented model. In addition, the validation in similar flow conditions shows a certain extent of generalization ability.

---

## 1. Introduction

The emergence and development of turbulent flow separation is crucial to the performance of a wing or turbomachinery. For example, trailing edge stall, which causes loss of lift coefficient, is induced by the breakup of the separation bubble appearing from the trailing edge [1]. In contrast, separation on an airfoil with horn ice usually begins at the leading edge at a small angle [2], which is characterized by the abrupt emergence of a separation bubble behind the ice shape and vast loss of the lift coefficient.

The numerical simulation of turbulent flow separation is essential for aerodynamic design applications. Although high-fidelity methods such as direct numerical simulation (DNS) and large eddy simulations (LES) can produce reliable results, the computational cost is too high to apply to daily engineering design applications. Efficient and reliable Reynolds-averaged Navier-Stokes (RANS) models will continue to be applied in daily engineering design in the foreseeable future [3]. The consensus is that the turbulence model is the most important factor to accurately evaluate the behavior of flow separation. Nevertheless, widely used RANS models such as the one-equation Spalart–Allmaras (SA) model and two-equation $k - \omega$ shear stress transport (SST) model can fail to make accurate predictions in the presence of flow separation. Studies show that the results of SA and SST models are far from satisfied in predicting the clean airfoil stall performance caused by boundary layer separation [4] or the iced airfoil stall performance induced by separating shear layer [2].

Previous studies have revealed that the nonequilibrium effect in the separating shear layer is one of the reasons for the failure of the present RANS models. Most turbulence models are calibrated by equilibrium turbulent shear flow, where the production-to-dissipation rate of turbulence kinetic energy is approximately 1.4 to 1.8 [5]. However, experimental and numerical studies have shown that the production of turbulence kinetic energy could be significantly larger than dissipation in a separating shear layer [6] or in the region of corner separation, as well as the wake and boundary layer of a compressor flow [7]. There have been some modifications for the turbulence models to include the nonequilibrium effect in a separated flow. In our previous studies, Li et al developed a separating shear layer fixed (SPF) $k - \overline{v^2} - \omega$ model by applying a multiplier to the destruction term of the $\omega$ equation to improve the accuracy of stall prediction for an iced airfoil [8][9][10].

However, ad hoc modifications for turbulence modeling usually have poor generality on different flow problems. For example, Rumsey [11] found that although the enhancement of turbulent mixing in a separated shear layer improves the prediction accuracy for a periodic hill, the same corrections obtained worse results for backstep flows. This indicates that the nonequilibrium turbulence behavior might be different in different flow cases, which is difficult to precisely describe by manual functional corrections. On the other hand, the previous correction of Li et al [8] merely increased the production-to-dissipation rate in the shear layer region, while the LES results by Fang et al [7] showed that the production can be either increased or decreased in such regions. Consequently, a more rigorous process of quantifying and reconstructing model corrections, especially for the nonequilibrium effect of the separated flow, is needed.

In recent years, data-driven approaches have been increasingly applied to turbulence modeling. This benefits from the increasing data from experiments or high-fidelity simulations and the application of statistical and machine-learning methods in turbulence modeling. From the aspect of data-driven turbulence modeling, the simplifications for Reynolds stress closure bring uncertainty to the RANS simulation. The uncertainty propagates forward through RANS computations and ultimately leads to uncertainty in the predictions. Data-driven methods can be used for quantifying or reducing the uncertainties of the RANS models [12] and can be further used for the augmentation of RANS models.

There are typically two types of studies focused on the uncertain model parameter $\theta$ and its prior probability distribution $p(\theta)$. From the forward perspective, the probability distribution of QoI can be determined by sampling from $p(\theta)$ and computing QoIs for each sample [12][13]. From the backward perspective, when given observable data $D$ from DNS or experiments, the posterior probability distribution of model parameters $p(\theta|D)$ can be inferenced by the Bayes formula:

$$p(\theta|D) \propto p(D|\theta)p(\theta) \tag{1}$$

Markov chain Monte Carlo (MCMC) methods [13][14] or ensemble Kalman filtering (EnKF) [15][16] are commonly used to obtain the precise or approximate posterior probability distribution. Maximum a posteriori (MAP) inference can be used if only the optimal values of $\theta$ are needed. This is also called field inversion or data assimilation.

Field inversion has been applied in many flow problems to correct either the model parameters or the model-form uncertainties. Papadimitriou et al. used the high-order adjoint method to infer the coefficients of the SA model [16]. Oliver et al. used Bayesian inference to quantify the uncertainty of model parameters and stochastic extensions of several eddy viscosity turbulence models [17]. Dow and Wang used the adjoint method to infer the viscosity field in a plain channel flow to fit the velocity field obtained from DNS [17]. Singh and Duraisamy studied the flow around an airfoil at a high angle of attack, where a strong adverse pressure gradient and flow separation existed [18]. They modified the SA model by multiplying the production term by a factor $\beta(x)$. Foures et al. used the continuous adjoint method to infer the Reynolds stress directly [19]. Zhang et al. used constraints based on physics knowledge for the field inversion of channel flow [20]. Apart from steady flow problems, data assimilation has also been used for unsteady problems. Meldi and Poux proposed the sequential data assimilation method using Kalman filtering [21], and He et al. improved the method using the continuous adjoint method [22]. Field inversion is usually converted to an optimization problem based on the adjoint method [23]. However, there may be innumerable local optimal solutions for the optimization problem. Many of them are physically unfeasible and may be obtained by gradient-based optimization algorithms. It is a critical problem to consider the physical mechanism in the field inversion process and locate the model uncertainties with physical realizability. Setting the prior based on physical intuitions has been proven to be an effective approach for this problem [16][20].

Machine learning (ML) is another frequently used approach for data-driven turbulence modeling. Many studies have used ML models to assist RANS modeling by predicting the source terms in model equations [18], eddy viscosity [24] or Reynolds stress [25][26] during CFD iterations. The model predictions can be reproduced using

ML techniques without knowledge of the functional form of the original model terms [18], verifying the feasibility of reconstructing or enhancing existing turbulence models. Consequently, after determining the spatially distributed model corrections by field inversion, a functional relationship between local flow features and model corrections can be established using ML to extract generalized modeling knowledge. Then, an enhanced turbulence model can be obtained using the ML model to generalize the model corrections. Parish and Duraisamy put forward field inversion and machine learning (FIML) as a new paradigm for data-driven predictive modeling [23][27]. They applied FIML to a scalar ordinary differential equation and turbulent channel flow. The high-order discrete adjoint method is used to obtain the model corrections, and Gaussian processes are used for ML [23]. Singh et al. also used an artificial neural network (ANN) to obtain an enhanced SA model, which reproduces the experimental $C_L$ for separation flow around an airfoil [28]. Ferrero et al. used similar methods to obtain a data-augmented SA model for turbomachinery flows [29]. They used the wall isentropic Mach number distribution as the experimental data for correction, and they limited the range of correction to improve the robustness of the correction factor. Previous studies have established FIML as an effective framework for the correction of RANS models. However, how to consider the physical mechanism in field inversion is a problem that has drawn little attention. For example, nonequilibrium turbulence is an important physical mechanism of a separated shear layer. Constraining the correction area and limiting the correction range in the field inversion based on the nonequilibrium turbulence phenomenon might be an effective method to obtain a physically feasible modification of turbulence modeling.

In this paper, FIML is used for data-driven turbulence modeling considering the nonequilibrium effect. The SA model is modified by multiplying the production term of $\tilde{\nu}_t$ with a spatially distributed correction factor $\beta(x)$, which is closely related to the equilibrium state of the turbulence. The corrections are applied to several problems, namely, the S809 airfoil, a periodic hill, and the GLC305 airfoil with ice shape 944, to quantify the model uncertainty caused by the nonequilibrium effect. The discrete adjoint method is used for the inverse problem with limited experimental data. The prior correction is determined considering physical knowledge, and the results show good consistency with previous studies by Li et al. [8][9][10]. Then, an ANN is used to learn the functional map from local flow features to the correction. An augmented SA model is built by embedding the ANN into the SA model to predict the correction multiplier $\beta(x)$. The augmented SA model is validated with the same flow conditions as the datasets for field inversion, and the experimental QoIs are successfully reproduced. Finally, the augmented model is examined with flow conditions similar to those of the datasets, and a certain generalization capability is demonstrated.

## 2. Framework of field inversion and machine learning

The framework used in this paper is similar to that in references [28][29], as shown in Fig. 1. First, field inversion is implemented in different flow cases to obtain corresponding model corrections of the SA model. Then, the dataset from field inversion is used to train an ANN. Finally, the ANN is integrated into the original SA model to form an ML augmented model.

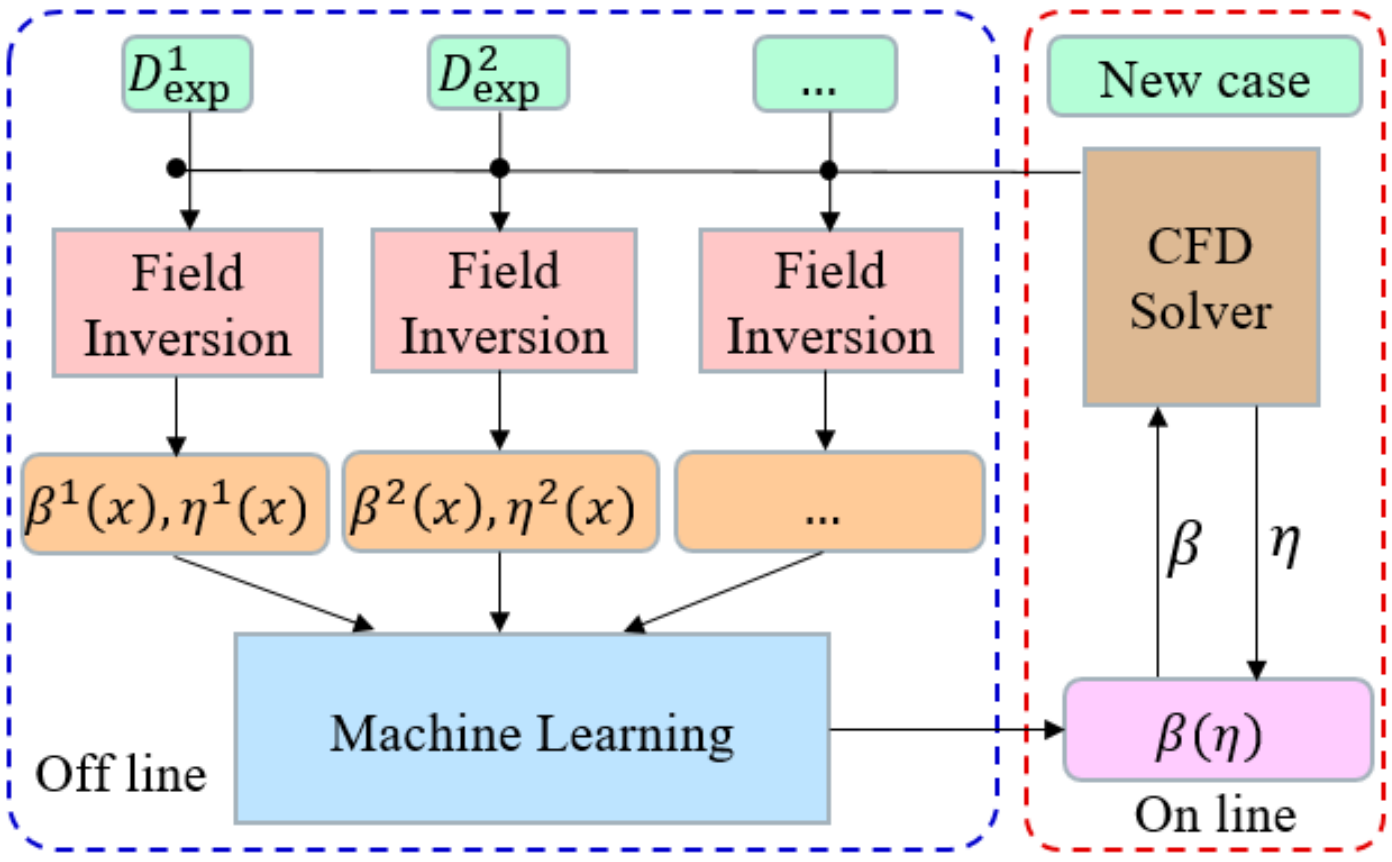

Fig. 1 Framework of FIML used in this paper

## 2.1 *Field inversion*

The SA model [30] contains a transport equation for eddy viscosity $\tilde{\nu}_t$, which is written as:

$$\frac{\mathrm{D}\tilde{\nu}}{\mathrm{D}t} = P - D + \frac{1}{\sigma}\left[\nabla \cdot \left((\nu + \tilde{\nu})\nabla\tilde{\nu}\right) + C_{b2}(\nabla\tilde{\nu})^2\right] \tag{2}$$

Here, $\nu$ is the molecular viscosity. $\sigma = 2/3$, $C_{b2}$ is a model parameter, and $\tilde{\nu}$ is the solution variable, which is used to compute the eddy viscosity $\nu_t$:

$$\nu_t = \tilde{\nu} f_{v1},\ f_{v1} = \frac{\chi^3}{C_{v1}^3 + \chi^3},\ \chi = \frac{\tilde{\nu}}{\nu} \tag{3}$$

$P$ is the production term, and $D$ is the destruction term:

$$P = C_{b1}\tilde{S}\tilde{\nu},\ D = C_{w1} f_w \left[\frac{\tilde{\nu}}{d}\right]$$

$$\tilde{S} = (1 - f_{t2})\|\boldsymbol{S}\| + \frac{\tilde{\nu}}{\kappa^2 d^2}[(1 - f_{t2})f_{v2} + f_{t2}]$$

$$f_w = g\left(\frac{1 + C_{w3}^6}{g^6 + C_{w3}^6}\right)^{\frac{1}{6}}$$

$$g = r + C_{w2}(r^6 - r)$$

$$r = min\left(\frac{\tilde{\nu}}{\kappa^2 d^2 \hat{S}}, 10.0\right)$$

$$\hat{S} = \max\left(\|S\| + \frac{\tilde{\nu} f_{v2}}{\kappa^2 d^2}, 0\right)$$

$$f_{v2} = 1 - \frac{\chi}{1 + f_{v1}\chi}$$

$$f_{t2} = C_{t3} e^{-C_{t4}\chi^2} \tag{4}$$

where $\boldsymbol{S}$ is the strain rate tensor of the mean velocity field, $d$ is the minimum distance to the wall, $\kappa = 0.41$, and $C_{b1}, C_{w1}, C_{w2}, C_{w3}, C_{v1}, C_{t3}$ and $C_{t4}$ are model constants.

Although the SA model has been validated in many flow cases, there are several flow problems that the model has poor accuracy, such as the massive flow separation problem. Many efforts have been made to improve the SA model for flow separation problems by implementing modifications to the model terms, such as SAE [31] and SA-QCR [32]. In this paper, a data-driven augmentation approach is applied to the SA model by enhancing its model parameters. The production term of $\tilde{\nu}$ is multiplied by a spatially distributed factor $\beta$, making the transport equation:

$$\frac{\mathrm{D}\tilde{\nu}}{\mathrm{D}t} = \beta \cdot P - D + \frac{1}{\sigma}\left[\boldsymbol{\nabla} \cdot \left((\nu + \tilde{\nu})\boldsymbol{\nabla}\tilde{\nu}\right) + C_{b2}(\boldsymbol{\nabla}\tilde{\nu})^2\right] \tag{5}$$

Statistically speaking, rather than physical analysis, $\beta(\boldsymbol{x})$ might be a random field. The prior probability distribution $\mathrm{p}(\boldsymbol{\beta})$ can be determined manually with prior knowledge of turbulence modeling before any observation data $\boldsymbol{d}$ from experiments or DNS is given. Once given observation data $\boldsymbol{d}$, the posterior probability distribution of $\boldsymbol{\beta}$ can be obtained by Bayesian inference. We assume that $\boldsymbol{\beta}$ and $\boldsymbol{d}$ are Gaussian and independent of each other on different grid points while having the same variance. Furthermore, supposing a prior of 1.0 for $\beta$ in each grid point, the posterior of $\boldsymbol{\beta}$ can be written as:

$$J = \sum_i \sigma_{obs}^{-2} \left(d_i - h_i(\boldsymbol{\beta})\right)^2 + \sum_j \sigma_{\beta}^{-2} \left(\beta_j - 1\right)^2 \tag{6}$$

where $\sigma_{obs}^2$ and $\sigma_{\beta}^2$ are the variance for $\boldsymbol{d}$ and $\boldsymbol{\beta}$, respectively, and $h_i(\boldsymbol{\beta})$ is the predicted QoI at the i-th observed point with a given $\boldsymbol{\beta}$. In this paper, MAP inference is used to find the optimal correction $\boldsymbol{\beta}$, which is equivalent to an optimization problem with an object function $J$. The first part of $J$ represents the prediction error with a given $\boldsymbol{\beta}$, and the second part of $J$ represents the deviation from the prior of $\boldsymbol{\beta}$, which can also be interpreted as regularization. The weight between these two terms is adjusted by tuning $\sigma_{obs}$ and $\sigma_{\beta}$.

### *2.2 Discrete adjoint method*

The adjoint method is an effective approach to compute the derivative of an object function with respect to a large set of variables when involving partial differentiation equations (PDEs). This method is often combined with gradient-based optimization algorithms to solve optimization problems [33][34]. Due to its low computational cost to the number of variables, the method is particularly suitable for the optimization problem in field inversion, where the parameter on each grid point might be a design variable. The discrete adjoint method [35][36] is used in this paper. The algorithm is briefly introduced as follows.

The aerodynamic object function $J$ and the residual of RANS equations $\boldsymbol{R}$ can be expressed as functions of the coordinate $\boldsymbol{x}$ and the flow variables $\boldsymbol{w}$ at each grid point: $J(\boldsymbol{w},\boldsymbol{x})$, $\boldsymbol{R}(\boldsymbol{w},\boldsymbol{x})$. Given the constraint of RANS equations $\boldsymbol{R}(\boldsymbol{w},\boldsymbol{x}) = \boldsymbol{0}$, the derivative of $J$ with respect to the design variables $\boldsymbol{x}$ can be expressed as follows:

$$\boldsymbol{G} = \frac{dJ}{dx} = \frac{\partial J}{\partial x} + \frac{\partial J}{\partial w}\frac{\partial w}{\partial x} \tag{7}$$

If we take the differential of equation $\boldsymbol{R}(\boldsymbol{w},\boldsymbol{x}) = \boldsymbol{0}$, we obtain:

$$\boldsymbol{0} = \frac{d\boldsymbol{R}}{dx} = \frac{\partial \boldsymbol{R}}{\partial x} + \frac{\partial \boldsymbol{R}}{\partial w}\frac{\partial w}{\partial x} \tag{8}$$

Thus:

$$\boldsymbol{G} = \frac{\partial J}{\partial x} + \frac{\partial J}{\partial w}\frac{\partial w}{\partial x} = \frac{\partial J}{\partial x} - \frac{\partial J}{\partial w}\left(\frac{\partial \boldsymbol{R}}{\partial w}\right)^{-1}\frac{\partial \boldsymbol{R}}{\partial x} \tag{9}$$

We introduce an adjoint variable $\boldsymbol{\varphi}$**,** which satisfies:

$$\left[\frac{\partial \boldsymbol{R}}{\partial \boldsymbol{w}}\right]^{\boldsymbol{T}} \boldsymbol{\varphi} = -\left[\frac{\partial J}{\partial \boldsymbol{w}}\right]^{\boldsymbol{T}} \tag{10}$$

Then, we obtain:

$$\boldsymbol{G} = \frac{\partial J}{\partial x} + \boldsymbol{\varphi}^{T}\frac{\partial \boldsymbol{R}}{\partial x} \tag{11}$$

The discrete adjoint computation involves first obtaining $\frac{\partial \boldsymbol{R}}{\partial w}^{T}$ and $\frac{\partial J}{\partial w}$, solving the discrete adjoint equation to obtain $\boldsymbol{\varphi}$, then obtaining $\frac{\partial J}{\partial x}$ and $\frac{\partial \boldsymbol{R}}{\partial x}$, and finally obtaining $\boldsymbol{G}$ according to Eq. (11). The RANS solver and the discrete adjoint solver in this paper are based on the open-source CFD code ADFlow [37] with secondary development. The automatic differentiation code Tapanade [38] is used to generate the forward/reverse mode differentiation code for sensitivity evaluations.

In this paper, an augmented discrete adjoint method [39] is applied to handle the constraint of the aerodynamic optimization problem, such as the fixed lift coefficient $C_L$ of airfoil computation or specified mass flow rate of the internal flow.

*2.3 Machine learning*

Field inversion provides an optimal correction field $\boldsymbol{\beta}$ for each flow condition of the training datasets. This case-specific message can be generalized by establishing the functional relationship between the correction $\boldsymbol{\beta}$ and some local physics features. The choice of the features is crucial for the effect of the augmentation model and is usually dependent on physics intuition. The features used in this paper rely on [28] and [29]:

$$\left\{\chi, P, \frac{P}{D}, S, \frac{S}{\Omega}, \tau, f_d', |\nabla\tilde{\nu}|\right\}$$

where $\chi = \tilde{\nu}/\nu$ is the eddy viscosity ratio, $P$ and $D$ are respectively the production and destruction terms of the transport equation of $\tilde{\nu}$, $S$ and $\Omega$ are respectively the magnitudes of the strain rate and vorticity, $\tau$ is the magnitude of the Reynolds stress, and $|\nabla\tilde{\nu}|$ is the magnitude of the gradient of $\tilde{\nu}$.

The parameter $f_d'$ is proposed by [14] as a modification of $f_d$:

$$f_d' = 1 - tanh({r_d}^{0.5})$$

where $f_d = 1 - tanh((8r_d)^3)$ was initially proposed for detached eddy simulations [39] and was used as a feature for ML in reference [28].

Various ML algorithms have been used to assist RANS modeling. Parish et al. and Duraisamy et al. used the Gaussian process as the ML algorithm for FIML [23][27]. One advantage of Gaussian process regression is that it can predict both the mean and the variance of the output, so the uncertainty of the correction can be determined simultaneously. Wang et al. used a random forest to reconstruct the Reynolds stress model discrepancy [40]. Schmelzer used sparse symbolic regression to learn Reynolds stress expression [41], and Zhao et al used the gene-expression programming (GEP) method to train a symbolic turbulence model [42]. Singh et al. [28] and Ferrero et al. [29] used an ANN for FIML. Among those ML algorithms, the evaluation cost of ANNs is independent of the size of the training data. Because the ML model is called at each grid point and each timestep, the computational cost is an important factor for integration into an RANS solver. Consequently, an ANN is used as the ML algorithm in this paper. The structure of the ANN is shown in Fig. 2. Note that the specific structure of the ANN is chosen according to not only the error for regression but also the error for reproducing the experimental data when integrated into the SA model, and there might be no certain correlation between these two characteristics. This is discussed in the following section.

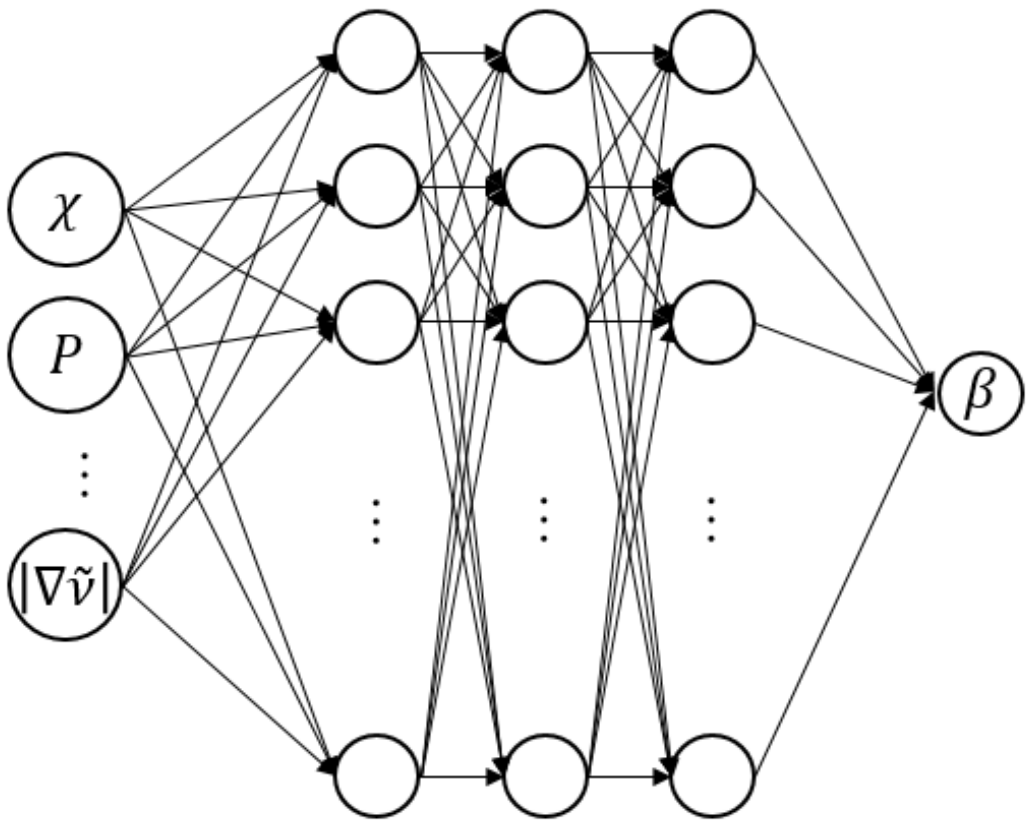

Fig. 2 Structure of the ANN used in this paper

## 3. Results and discussion

### *3.1 S809 airfoil stall case*

The S809 airfoil is designed for a wind-turbine blade with 21% relative thickness. The experimental data were obtained in the low-turbulence wind tunnel of the Delft University of Technology Low Speed Laboratory [43]. The flow around the S809 airfoil near the stall shows a strong adverse pressure gradient (APG), which causes separation at the tailing edge. Previous studies indicated that the SA model tends to overestimate the lift coefficient of the S809 airfoil near the stall [28]. The nonequilibrium turbulence in the shear layer might be one reason for the discrepancy [28]. In this paper, FIML is used to increase the accuracy of the SA model. Note that this case has been applied by many researchers [28][44]. It is also used as a validation case of the present FIML framework.

The computation grid is shown in Fig. 3. It has 457 points along the circumferential direction and 97 points in the normal direction. The height of the first grid layer is less than $1 \times 10^{-6}$c ($c$ is the chord length), which ensures $\Delta y^{+} < 1$.

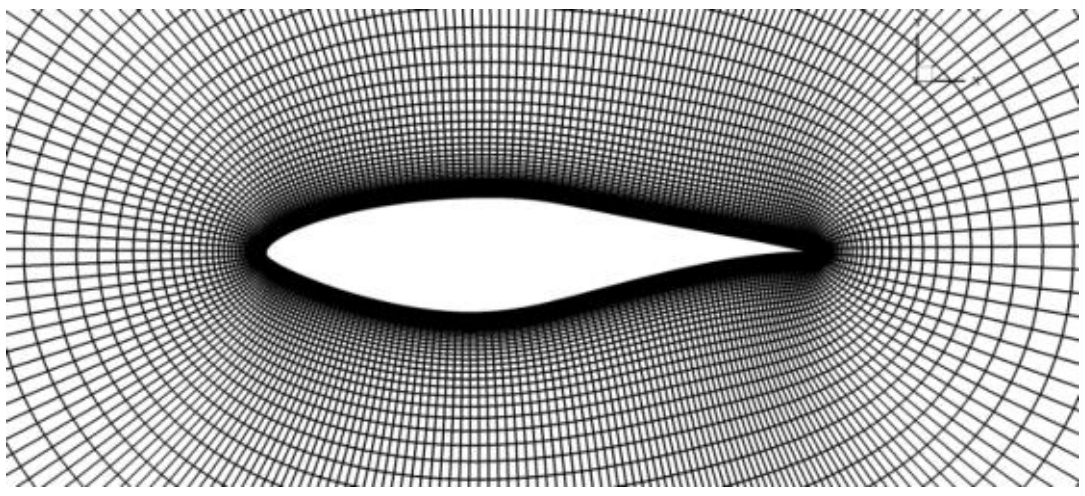

Fig. 3 Computation grid for the S809 airfoil

3.1.1 Field inversion of the S809 airfoil

The goal of field inversion is to obtain optimal correction fields for the SA model at large angles of attack, with which the aerodynamic force can match the experimental data. The field inversion is implemented under two flow conditions with $\mathrm{Ma} = 0.10, \mathrm{Re} = 2.0 \times 10^{6}$ and angles of attack at $\alpha = 8°$ and $14°$. The lift coefficient ($C_L$) from the experiment is chosen as the observation data for MAP inference. The object function is written as:

$$J = \sigma_{obs}^{-2}\left(C_L - h(\boldsymbol{\beta})\right)^2 + \sum_i \sigma_{\beta,i}^{-2}\left(\beta_i - 1\right)^2 \tag{14}$$

where $\sigma_{\beta,i}$ is the variance of the prior of $\beta$ at the $i$-th grid point, which creates a nonuniform variance field. For this case, $\sigma_{\beta,i}$ is set to be:

$$\begin{cases} \sigma_{\beta,i} = 200, & x<1.0 \\ \sigma_{\beta,i} = 20, & x>1.0 \end{cases} \tag{15}$$

This set of $\sigma_{\beta,i}$ comes from the physical knowledge that the uncertainty is mostly attributed to the strong APG near the boundary layer. A uniform variance field with $\sigma_{\beta,i} = 20$ is also used for field inversion for comparison. The convergence of $C_L$ and the object function $J$ at $\alpha = 14°$ at the two variance distributions are shown in Fig. 4. The error of the predicted $C_L$ is reduced sufficiently regardless of whether it has uniform variance or nonuniform variance. Fig. 5 shows the correction fields for the two variance distributions. The production of turbulence viscosity is suppressed mainly in the region with a strong APG near the boundary layer. On the other hand, the difference between the two correction fields indicates that there could be multiple solutions for the correction field, all of which can reduce the error of QoIs sufficiently. The introduction of physical knowledge by setting the prior can help exclude solutions that are inconsistent with physical expectations. Consequently, the nonuniform variance prior is used in the following to consider the physical mechanism of field inversion.

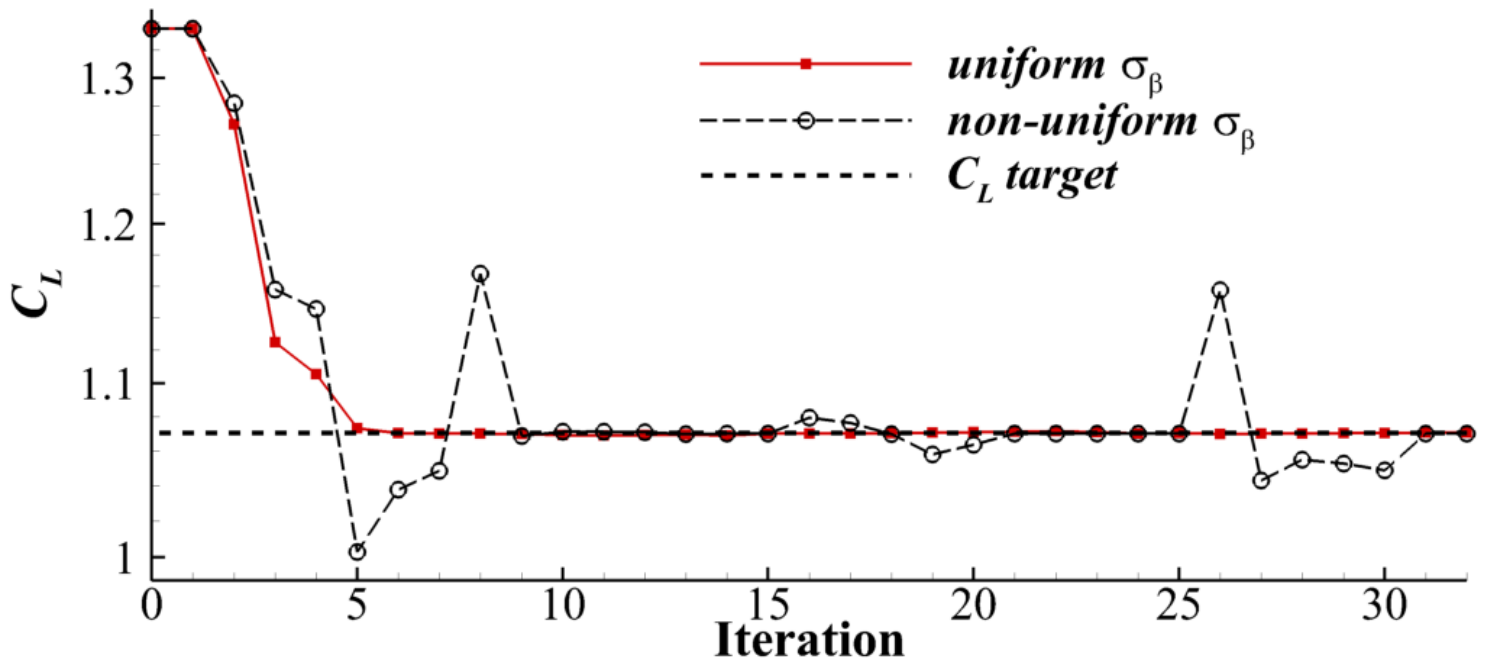

Fig. 4 Convergence of the squared error of $C_L$ at $\alpha = 14°$ for the two field inversions

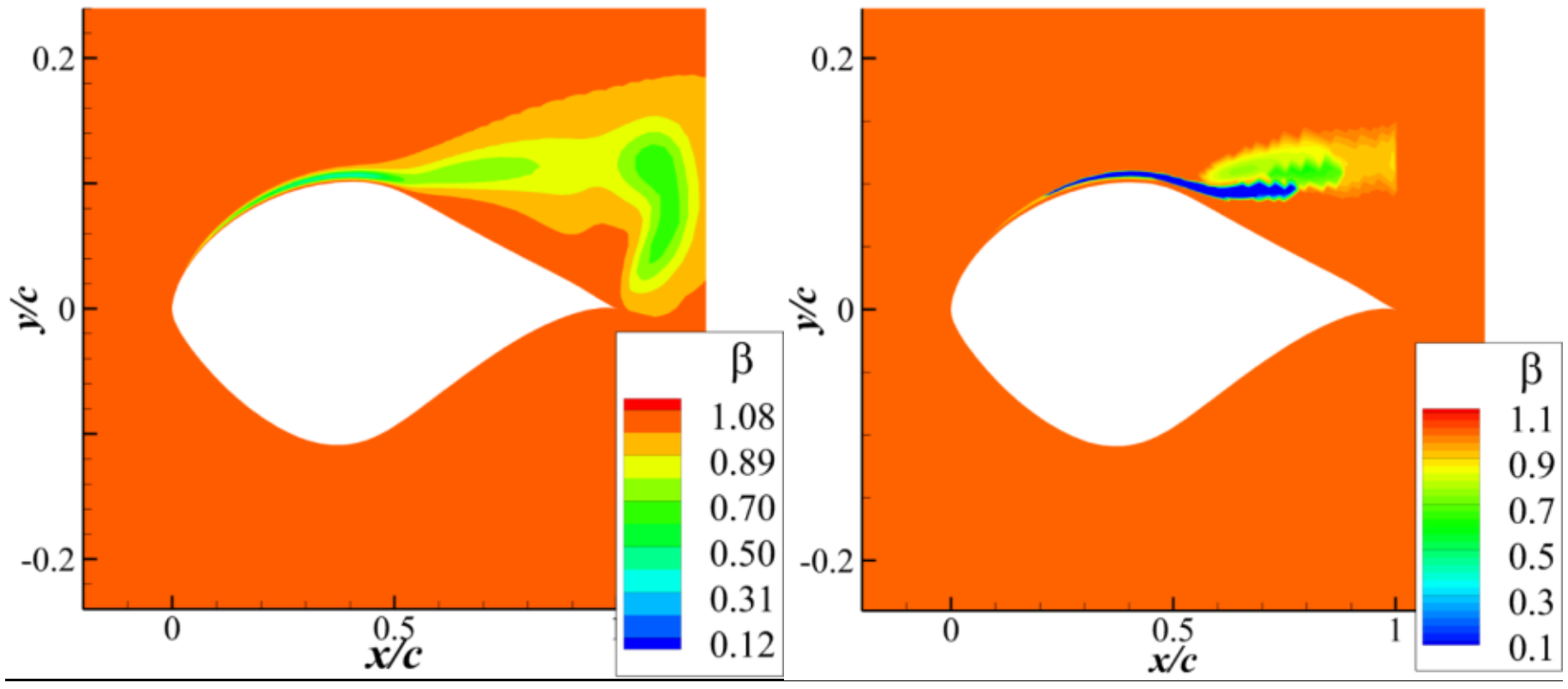

(1) uniform $\sigma_\beta$ (2) nonuniform $\sigma_\beta$

Fig. 5 Correction fields for the two field inversions

The flow fields predicted by the original SA model and field inversion at $\alpha = 8°$ and $14°$ are shown in Fig. 6. The separation bubbles are expanded after field inversion. The pressure distributions on the surface are compared with the experimental results in Fig. 7. Although the observation data involve only $C_L$, the accuracy of the $C_p$ distribution is significantly improved by field inversion. The results demonstrate the favorable effect of field inversion even with very limited data.

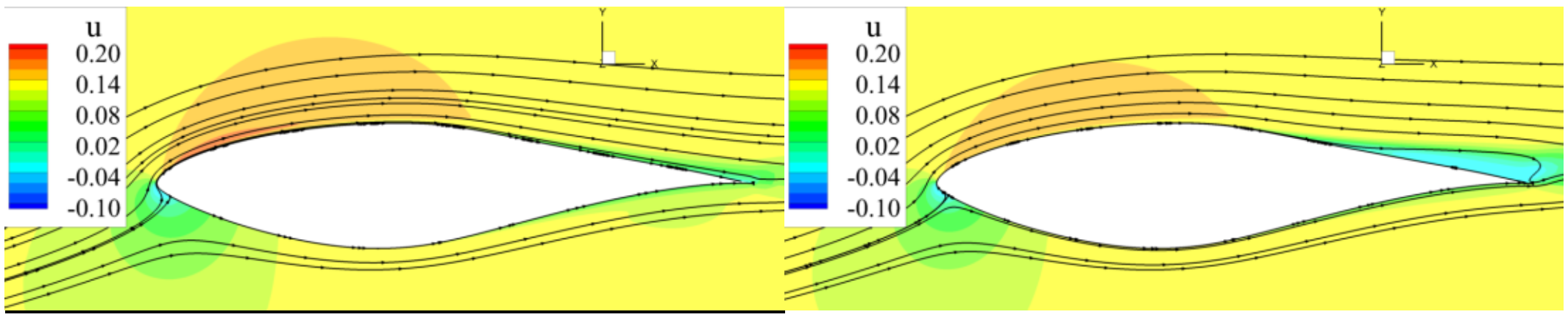

(1) $\alpha = 8°$, SA model (2) $\alpha = 8°$, field inversion

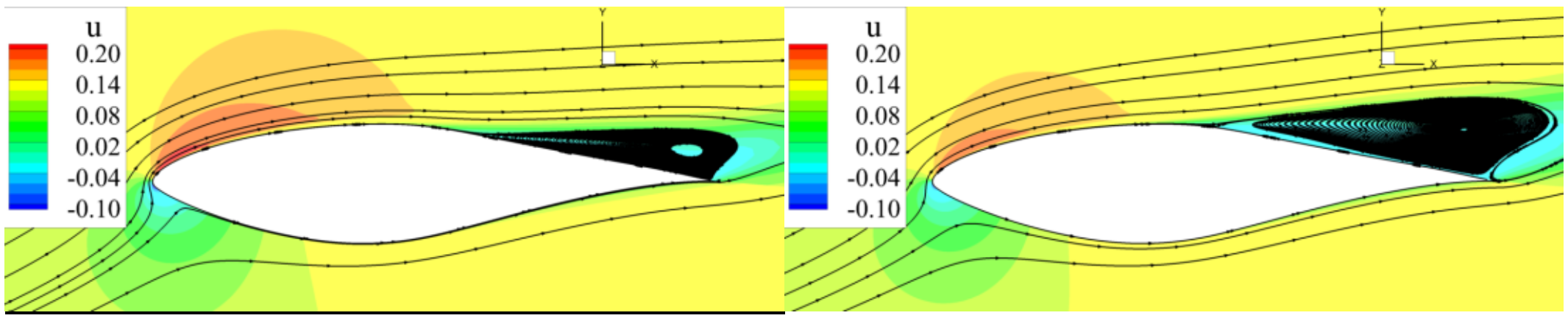

(3) $\alpha = 14°$, SA model (4) $\alpha = 14°$, field inversion

Fig. 6 Flow fields of the original SA model and field inversion at $\alpha = 8°$ and14°

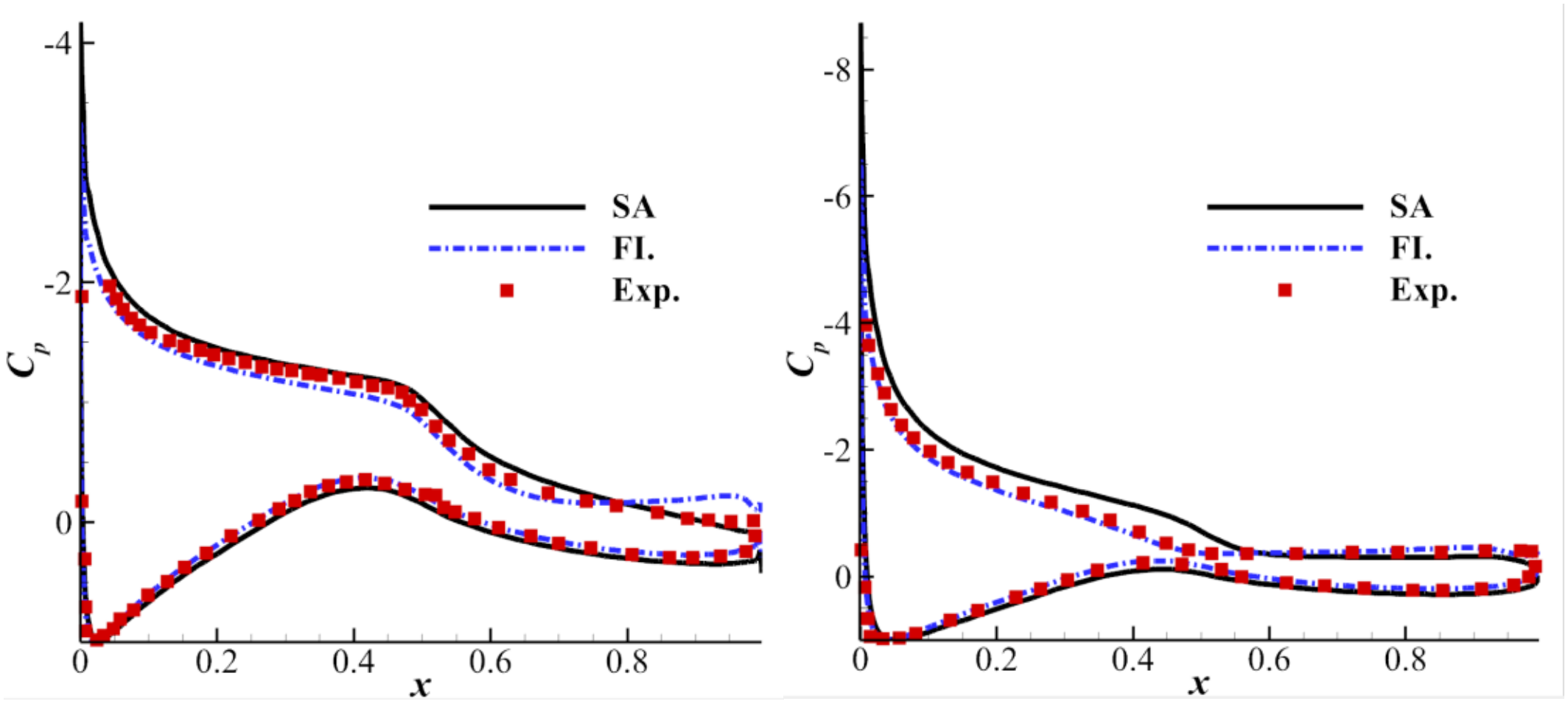

(1) $\alpha = 8°$ (2) $\alpha = 14°$

Fig. 7 Pressure distributions on the surface for the SA model and field inversion compared with experimental data

3.1.2 Machine learning based on the field inversion results

The field inversion results are used to train an ANN, which is further integrated into the SA model to enhance the prediction accuracy. There are 88658 samples in the dataset, including all grid points at both $\alpha = 8°$ and 14°. Fifteen percent of them are used for validation, while the rest are used for ANN training. The features of the field after field inversion at $\alpha = 14°$ are shown in Fig. 8. The open-source tool PyTorch [45] is used to train the ANN. The loss function is the mean square error (MSE), the training algorithm is Adam, and the activation function is sigmoid. Z-score normalization is used for the input features, and batch normalization is implemented between the hidden layers.

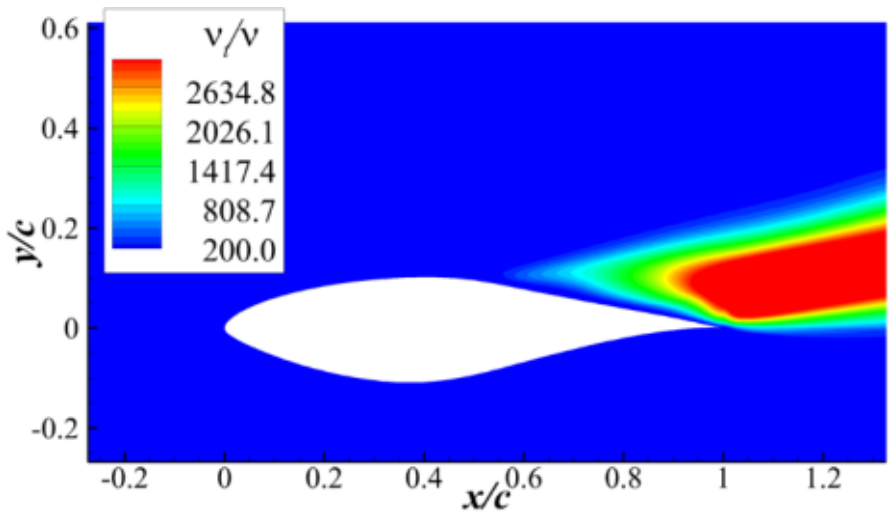

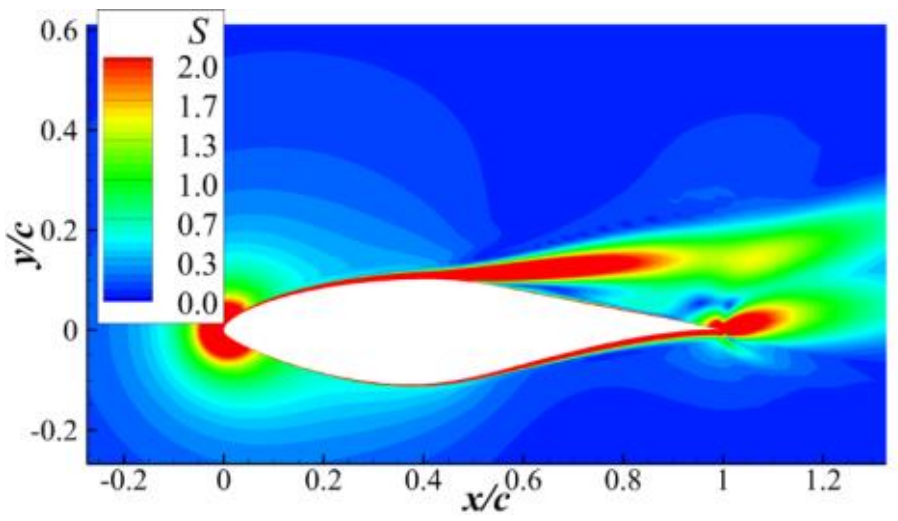

(1) $\frac{\tilde{\nu}}{\nu}$ (2) $S$

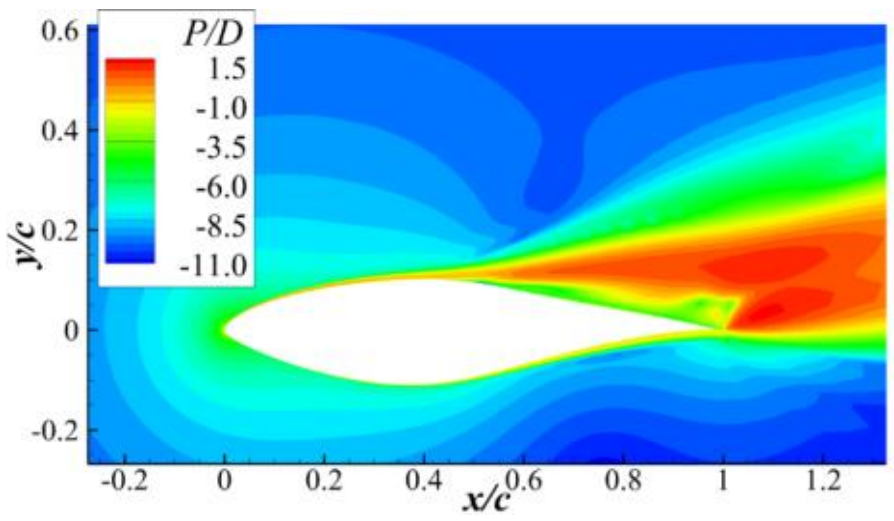

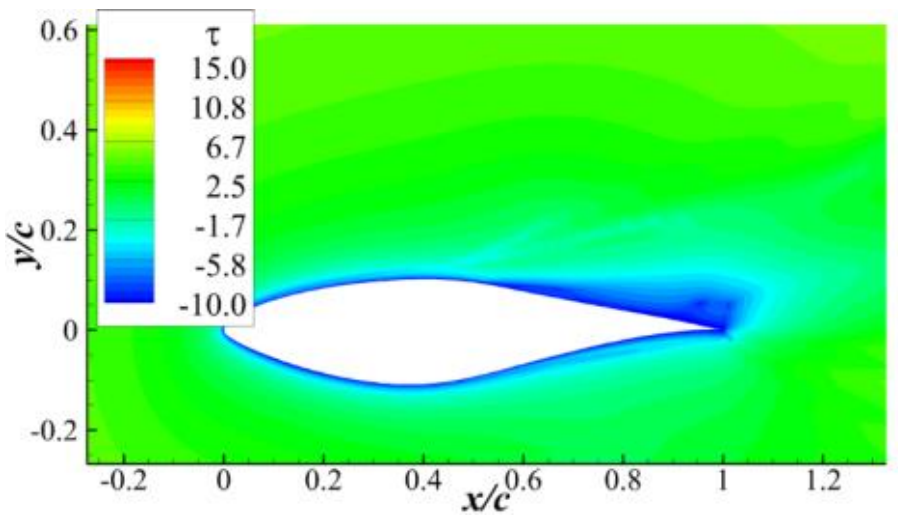

(3) $P/D$ (4) $\tau$

Fig. 8 Features from the field inversion at $\alpha = 14°$

The ANN has three hidden layers. Four ANNs with different sizes of hidden layers, $(8, 4, 2)$, $(8, 8, 8)$, $(16, 16, 8)$ and $(64, 32, 16)$, are tested with different training epochs. The root-mean-square errors on the training set and the validation set and the prediction errors of $C_L$ for the enhanced model are shown in Table 1.

Table 1 Root-mean-square errors on the training set and the validation set and the prediction error of $C_L$ for the enhanced model at $8°$ and $14°$

| Hidden layer settings | Number of epochs | Error in $\beta$ in training | | Error in $C_L$ in testing | |
|---|---|---|---|---|---|
| | | Training set | Validation set | $\alpha = 8°$ | $\alpha = 14°$ |
| $(8,4,2)$ | 100 | 0.072 | 0.067 | 9.0% | 2.3% |
| | 300 | 0.067 | 0.066 | 11.1% | -1.4% |
| | 500 | 0.068 | 0.064 | 15.1% | 3.9% |
| $(8,8,8)$ | 100 | 0.069 | 0.063 | 14.6% | 0.3% |
| | 300 | 0.062 | 0.061 | 14.8% | 4.4% |
| | 500 | 0.060 | 0.055 | 1.4% | -0.7% |
| $(16,16,8)$ | 100 | 0.064 | 0.071 | 2.8% | -6.6% |
| | 300 | 0.060 | 0.064 | **2.4**% | **-0.1**% |
| | 500 | 0.059 | 0.066 | 1.6% | 1.2% |
| $(64,32,16)$ | 100 | 0.061 | 0.066 | 1.7% | -4.1% |
| | 300 | 0.055 | 0.056 | 14.9% | 8.1% |
| | 500 | 0.068 | 0.064 | -2.0% | 2.5% |

The reduction in the loss function during training does not necessarily lead to the reduction in the prediction error for the enhanced SA model. Specifically, when the size of the hidden layer increases, the prediction error can be even larger while the loss function is still decreasing, which is a kind of overfitting. In fact, when the ANN is embedded into the SA model, the input features become different from the training datasets during iterations before convergence. The reduction in the loss function ensures only that the ANN makes correct predictions in the flow field after convergence; however, it does not ensure that the ANN makes suitable predictions to guide the iterations to a correct flow field before convergence. Different from an ordinary ML problem that determines the hyperparameters of the ANN by minimizing the loss function on the validation set, the optimal hyperparameters in this paper are determined based on the prediction error of the QoIs for the ANN-enhanced SA model.

The final size of the hidden layers of the ANN is $(16, 16, 8)$ with 300 training epocs. The distributions of $C_p$ computed by the enhanced SA model (FIML) compared with the baseline SA model and field inversion (FI) at $\alpha = 8°$ and $\alpha = 14°$ are shown in Fig. 9. It is shown that the enhanced SA model can reproduce the flow field from FI with high accuracy.

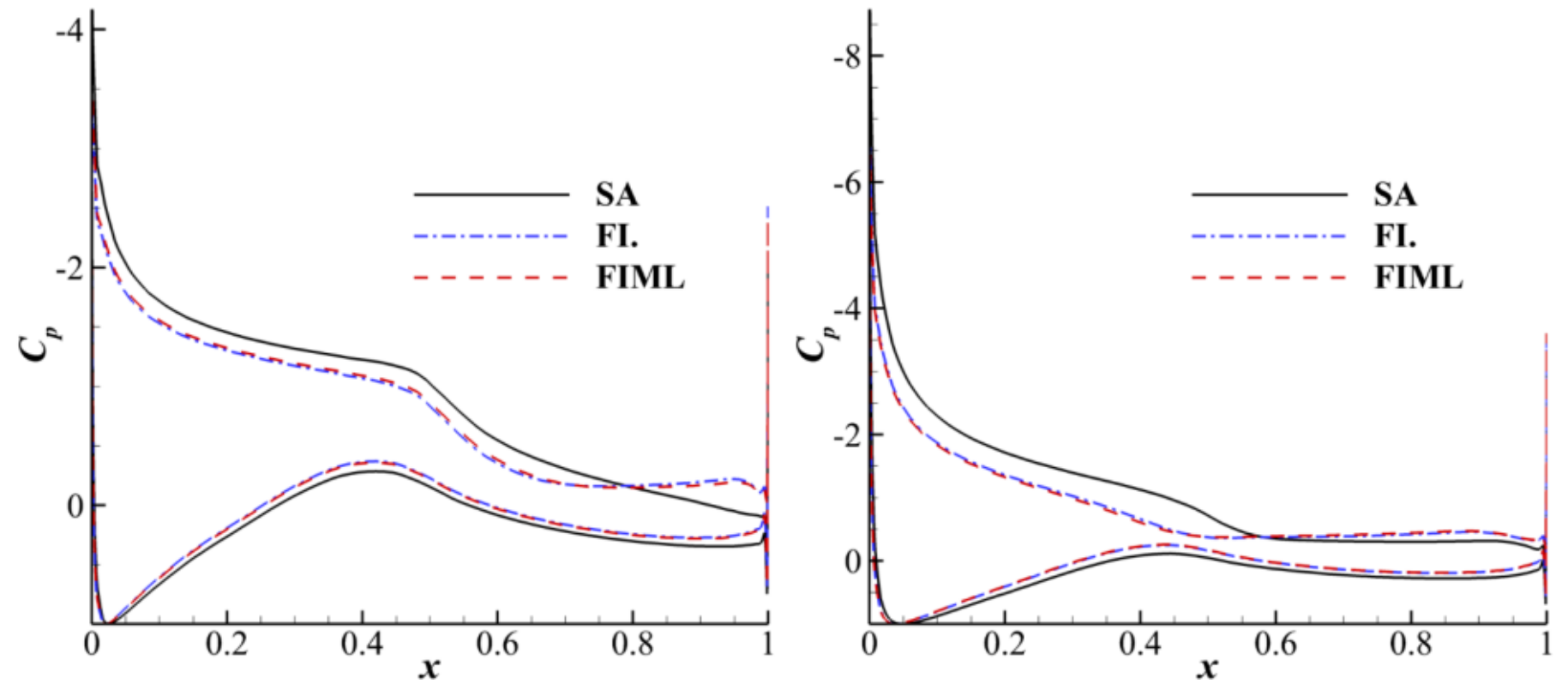

(1) $\alpha = 8°$ (2) $\alpha = 14°$

Fig. 9 Distribution of $C_p$ computed by the enhanced SA model compared with that of the original SA model and FI

To validate the generalization ability of the enhanced SA model, three different Reynolds numbers ($1 \times 10^6$, $2 \times 10^6$, and $3 \times 10^6$) are tested with a wide range of angles of attack. $C_L$ and $C_D$ computed by the original and enhanced SA model are compared with the experimental data in Fig. 10. The enhanced SA model shows considerable improvement not only in the two cases of FI but also in a wide range of Reynolds numbers and angles of attack. This indicates that the ML process has extracted generalizable modeling knowledge from the specific FI cases. Moreover, although the drag coefficient $C_D$ is not provided for the FI, the enhanced model could still make better predictions on $C_D$. This shows the powerful generalization ability of the FIML framework. The S809 airfoil case demonstrates that the present FIML framework is reliable.

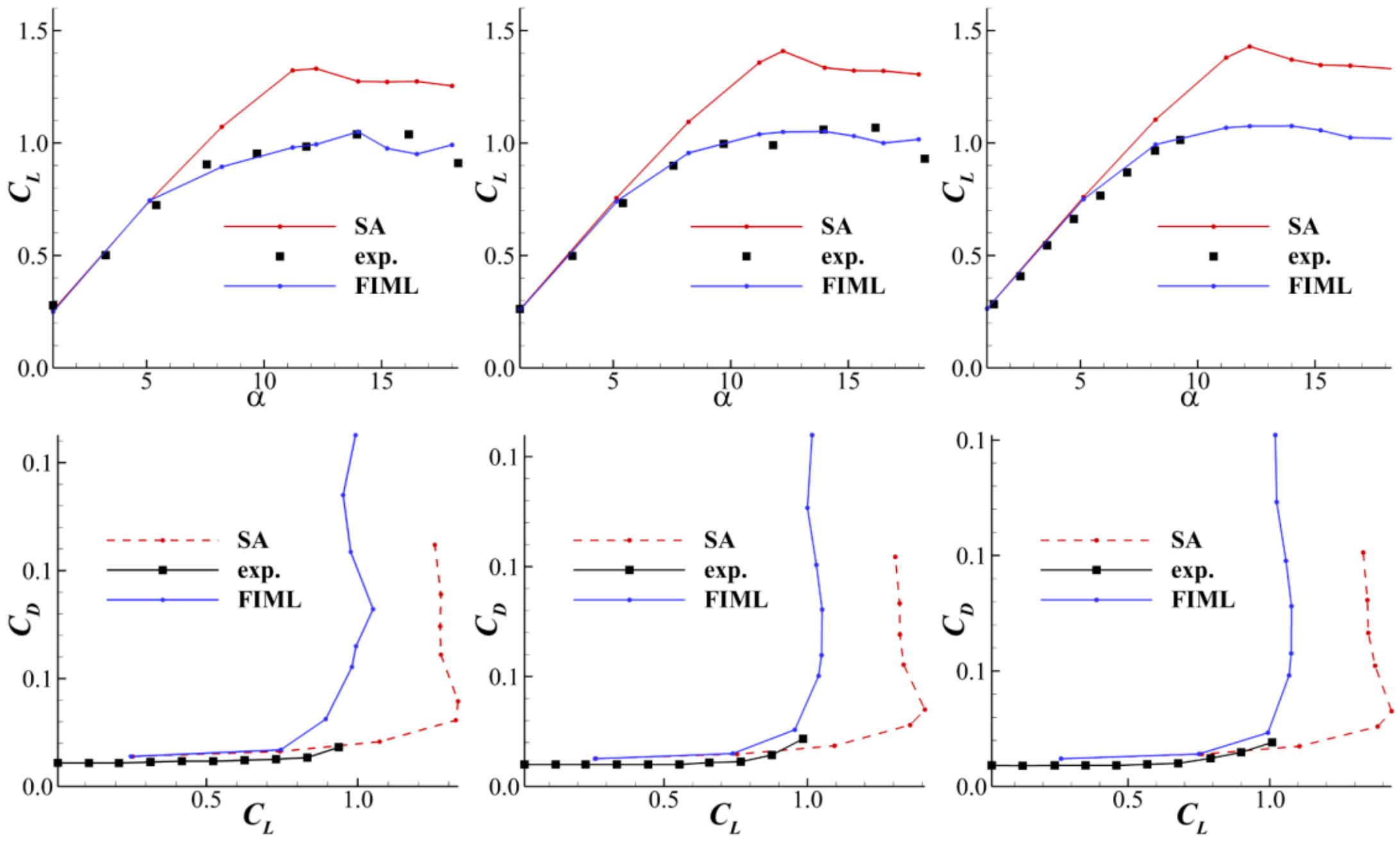

(1) $Re = 1 \times 10^6$ (2) $Re = 2 \times 10^6$ (3) $Re = 3 \times 10^6$

Fig. 10 $C_L$ and $C_D$ results from the original and enhanced SA model compared with experimental data at different Reynolds numbers

### 3.2 *Periodic hill case*

The periodic hill is a typical flow separation problem that is widely used for evaluating the performance of turbulence models. The large flow separation region makes this phenomenon challenging to simulate accurately. Extensive studies have focused on the correction of RANS models for periodic hill flows, using either manual turbulence modeling [11] or data-driven approaches [46][47]. It is known from physical analysis that the model uncertainty in this problem mostly comes from the nonequilibrium turbulence effect in the separation shear layer [11], where the turbulence shear stress can be underestimated. To fix the failure of turbulence models in separation regions, Rumsey [11] modified Menter's $k-\omega$ SST model. As shown in Fig. 11, they increased the destruction term of ω by a multiplier $F_{sf}$ in the separating shear layer. The fixed region is characterized by a nonequilibrium turbulence region, where the production-to-dissipation rate of turbulence kinetic energy is significantly larger than 1. With the same physics intuition, FIML is implemented on the periodic hill case to correct the original SA model in this paper.

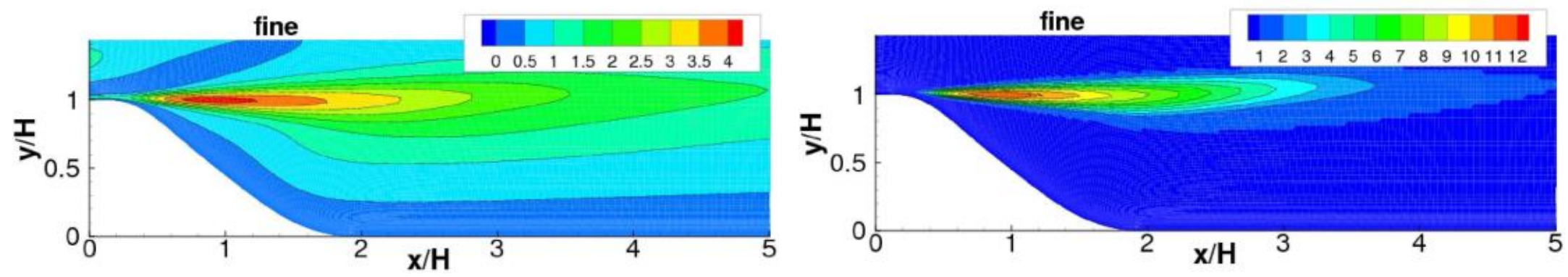

(1) $P/\varepsilon$ (2)$F_{sf}$

Fig. 11 Nonequilibrium turbulence effect and the corresponding modification in the separation shear layer of a periodic hill from Rumsey [11]

The geometries of the periodic hills used by Xiao et al. [48] are parameterized according to the steepness of the hill. The hill width is scaled by a factor $\alpha$ while the height remains fixed. The geometries with different $\alpha$ and the grid for case $\alpha = 0.8$ are shown in Fig. 12. There are 81 points in the streamwise direction and 77 points in the normal direction. The Reynolds number based on the crest height $b$ and the bulk flow velocity $v_b$ at the crest is $Re_b = 5600$. The upper and lower surfaces of the calculation domain adopt a no-slip wall boundary condition, and periodic conditions are applied at the inflow and outflow sections. The flow is driven by a uniformly distributed volume force. The volume force is automatically adjusted to match the bulk flow velocity $v_b$.

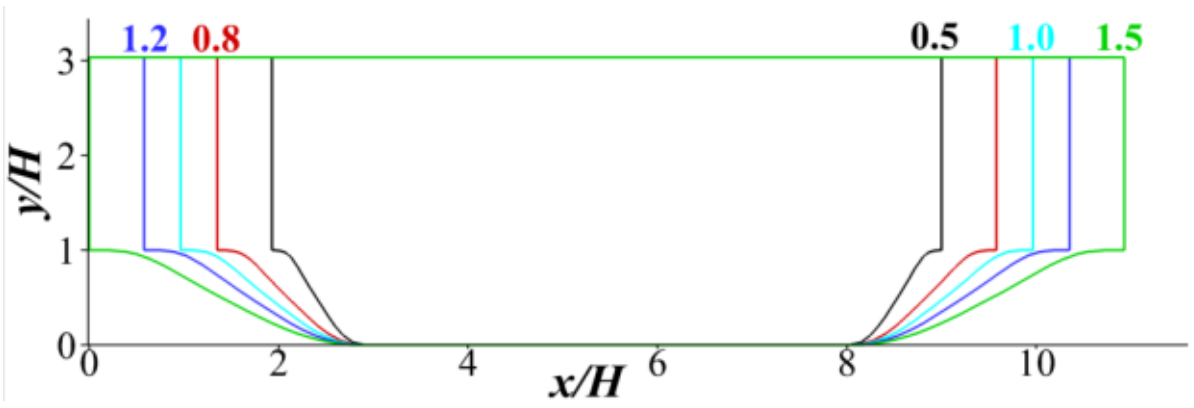

(1) Geometries with different $\alpha$

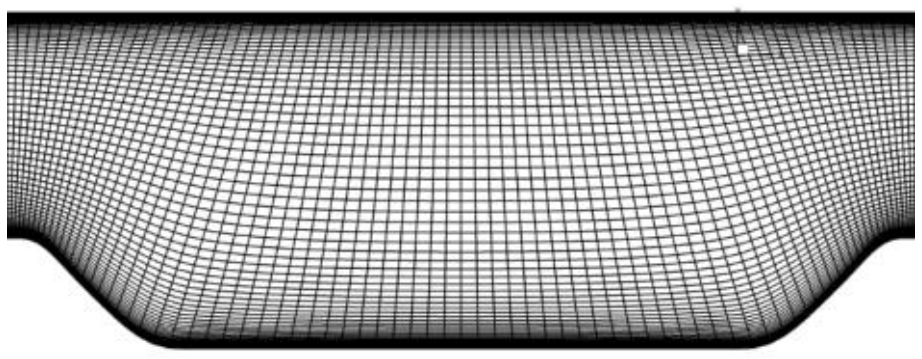

(2) Grid for the case $\alpha = 0.8$

Fig. 12 Geometry and computation grid for the periodic hill

3.2.1 Field inversion of the periodic hill

DNS results are obtained from reference [48]. The mean velocities at different locations from the DNS results are used as the observation data for FI. The locations are shown in Fig. 13. This leads to the following constrained optimization problem:

$$\min_{\boldsymbol{\beta}} J\,,\ J = \sum_i \left(u_i - h_i(\boldsymbol{\beta})\right)^2 + \sum_j \lambda_i \left(\beta_j - 1\right)^2, \text{s.t.}\, v_b = \text{const} \quad (17)$$

The constraint-augmented adjoint method is used for the sensitivity computation. The bulk flow velocity $v_b$ is used as the aerodynamic constraint, and the volume force in the field can be adjusted automatically.

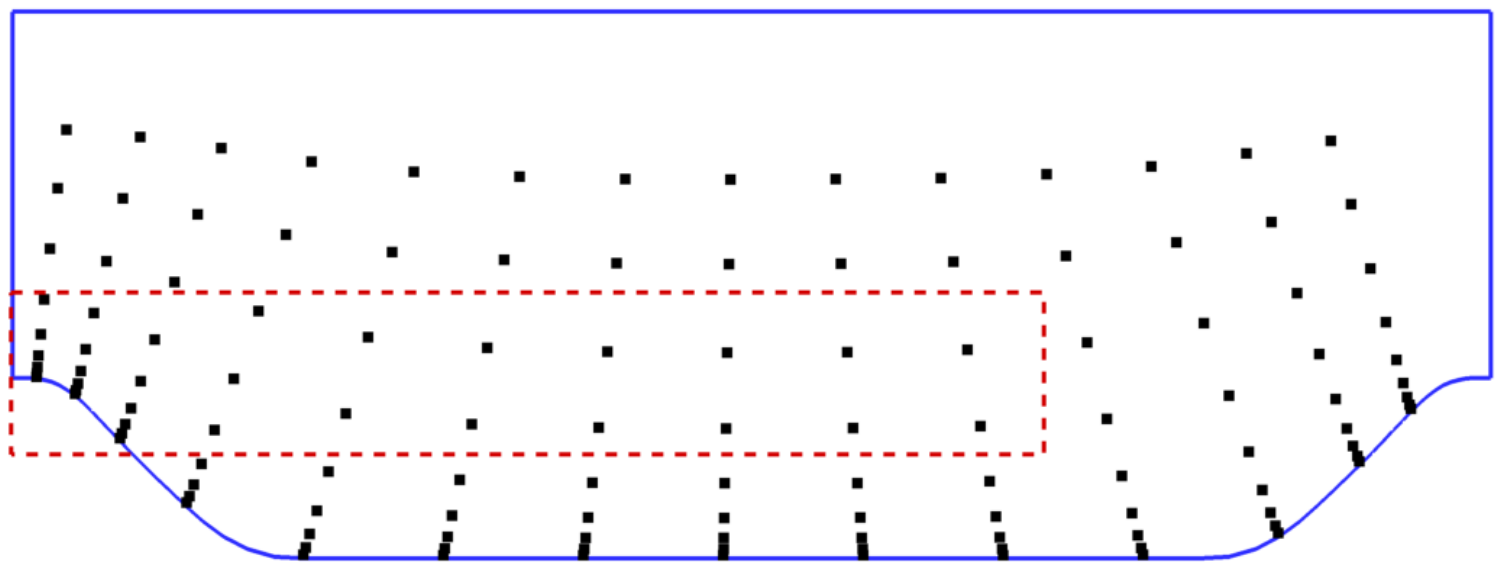

Fig. 13 Locations for mean velocity samples for FI

To make the inferenced correction field consistent with the physical analysis, regularization from prior knowledge is added in the optimization. The parameter $\lambda$ in Eq. (17) represent the weight of regularization. In general, a larger $\lambda$ reflects more confidence in the initial SA model. In this paper, three cases with different regularizations of $\boldsymbol{\beta}$ are made to study the influence of prior knowledge on the results of FI. The first case sets $\lambda_i = 0$, which uses no prior knowledge of $\boldsymbol{\beta}$. The second case limits the corrections in a prescribed rectangle region that covers the separation shear layer where the nonequilibrium turbulence effect is significant according to Rumsey [11]. The rectangular region is shown as the red dashed box in Fig. 13, where $\lambda = 8 \times 10^{-6}$ is set inside the region. The third case sets $\lambda = 8 \times 10^{-6}$ in the whole computation domain, which assumes a uniform variance. The setting for the three cases are listed in Table 2. The convergence histories of the squared error of velocity in different cases are shown in Fig. 14. The reduction in error is more difficult with a nonzero $\lambda$ than with $\lambda = 0$.

Table 2 Settings for the three cases

| Case number | Regularization factor $\lambda$ | Correction region |
|---|---|---|
| Case 1 | 0 | Whole flow field |
| Case 2 | $8 \times 10^{-6}$ | Inside the rectangular region |
| Case 3 | $8 \times 10^{-6}$ | Whole flow field |

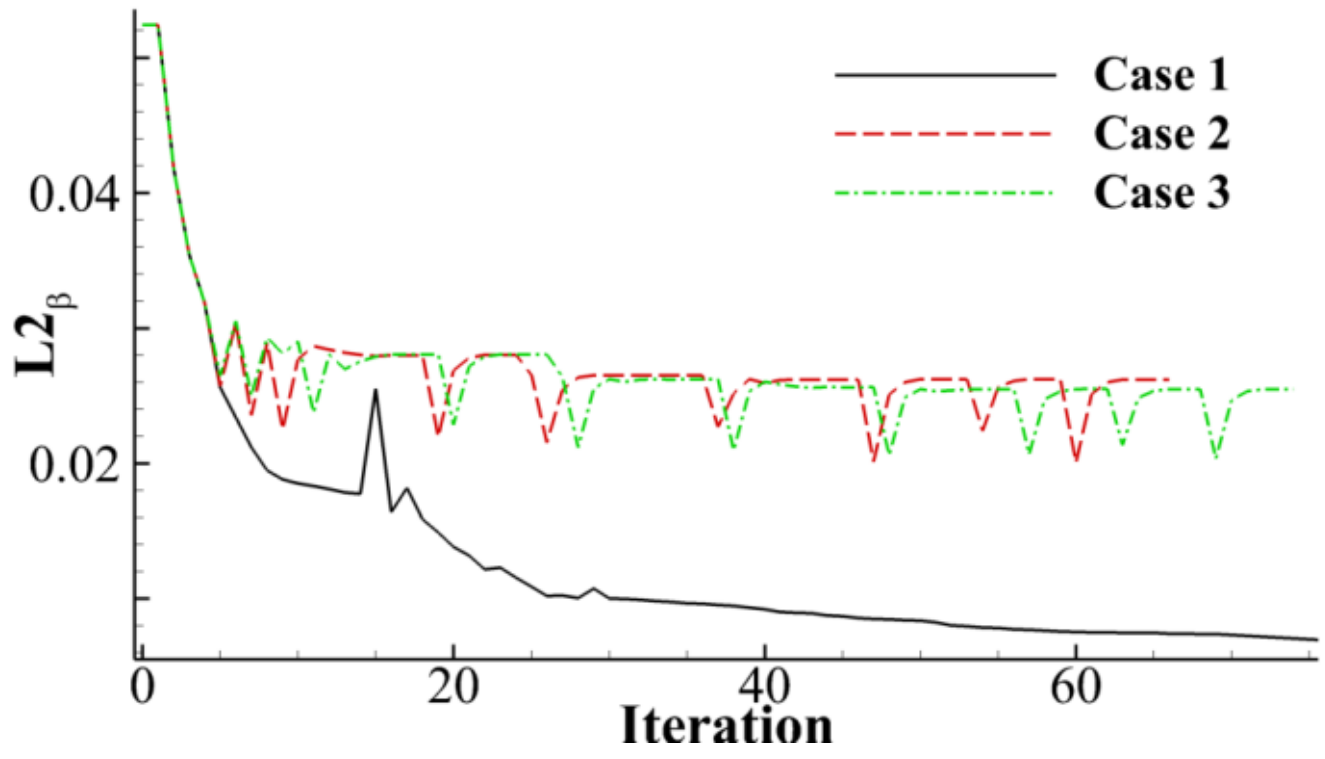

Fig. 14 Convergence histories of the error of QoIs in different cases

Fig. 15 compares the velocity profiles at different locations from FI with those of the original SA model and DNS results. The velocity distributions of the FI results are closer to the DNS data in all cases. The friction coefficients ($C_f$) are shown in Fig. 16. Although the friction coefficient on the lower surface is not explicitly included in the objective function, the FI results are also closer to the DNS results at most locations.

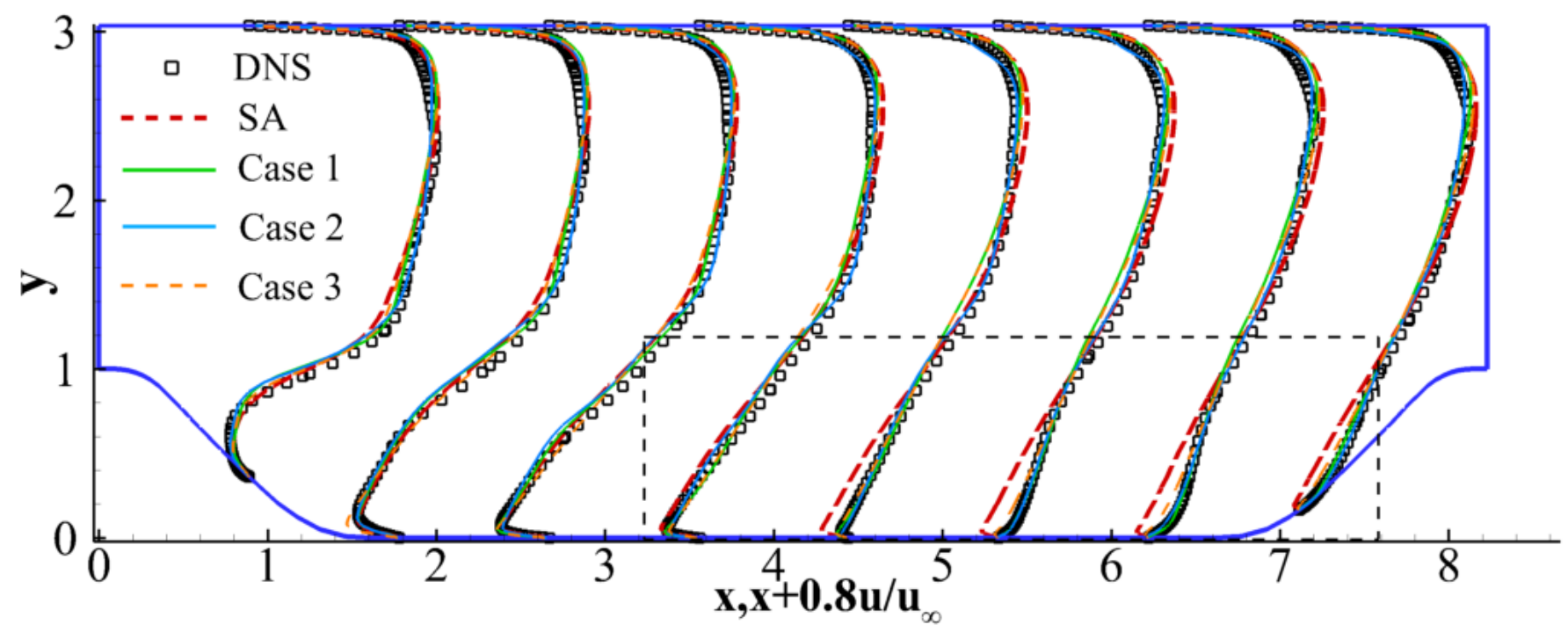

(1) Overall view of the velocity profiles

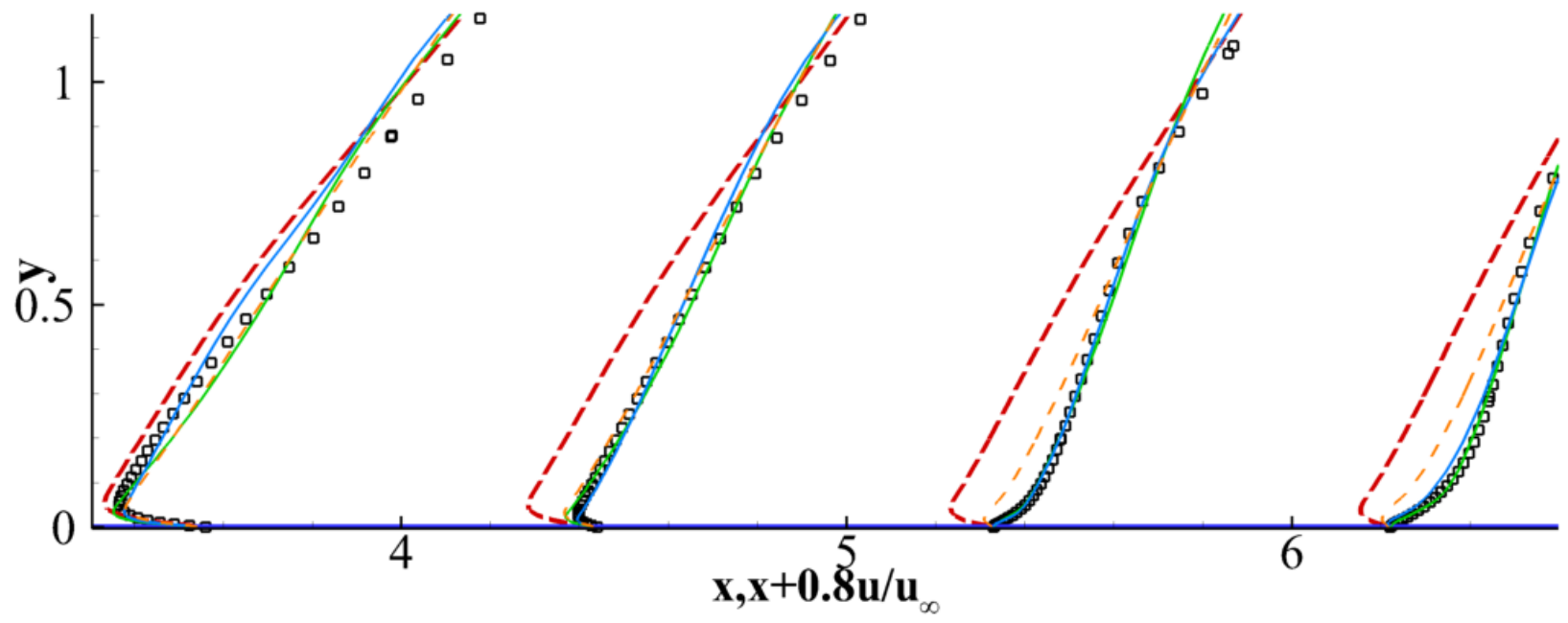

(2) Partial enlarged view of the dashed box in (1)

Fig. 15 Velocity profile at different locations from FI, original SA model and DNS results

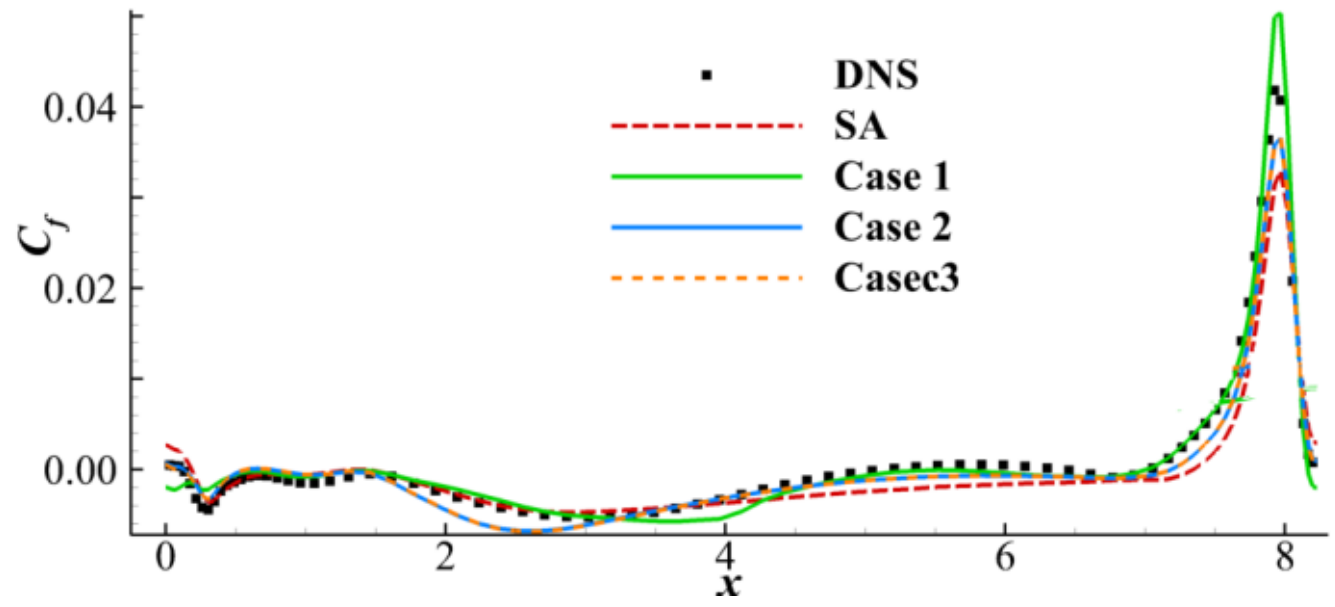

Fig. 16 Friction coefficient ($C_f$) on the lower surface from FI, the original SA model and DNS results

Fig. 17 shows the flow fields from the FI and DNS results. The locations of flow reattachment and the shapes of the separation bubble are well reproduced by FI. The separation point is located at the top of the hill, and the reattachment point is located at the middle bottom. The correction fields for the three cases are compared in Fig. 18. The correction field of the first case is very irregular, and no general tendency can be found from the field. A local optimized solution is obtained by the adjoint solver. However, the solution might be not physical. The second case constrains the corrections inside the prescribed region of the separation shear layer, which is marked as the dashed box in Fig. 18(2). The error of QoIs is still sufficiently reduced, with the production of $\tilde{\nu}$ decreasing in the starting location of the shear layer and enlarging downstream of the shear layer. This verifies the physics analysis that the model-form uncertainty mainly comes from the nonequilibrium turbulence effect inside the shear layer. The third case uses a uniform prior, which makes less human intervention for the FI. A similar correction is obtained as in the second case. This verifies the capability of FI to capture the model-form uncertainty in the critical region. Appropriate regularization is important to guide the FI to a physically realizable field. In the following section, the correction from the third case is chosen for the ML process.

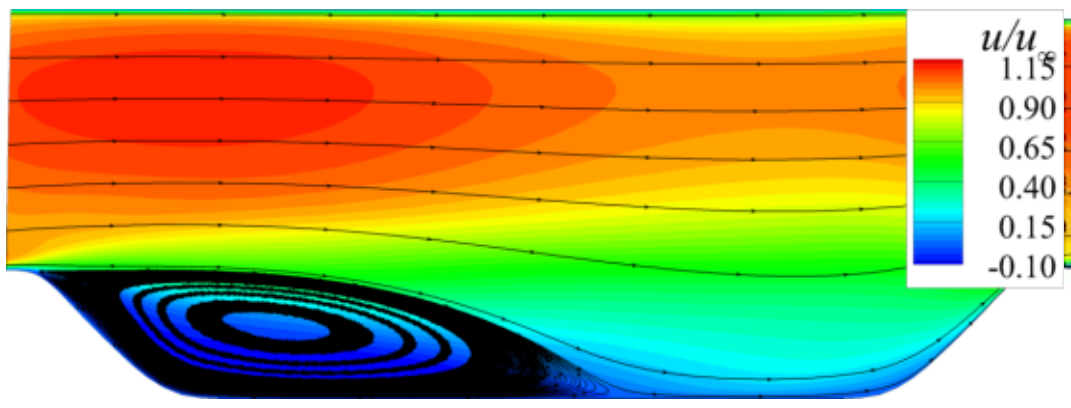

(1) DNS

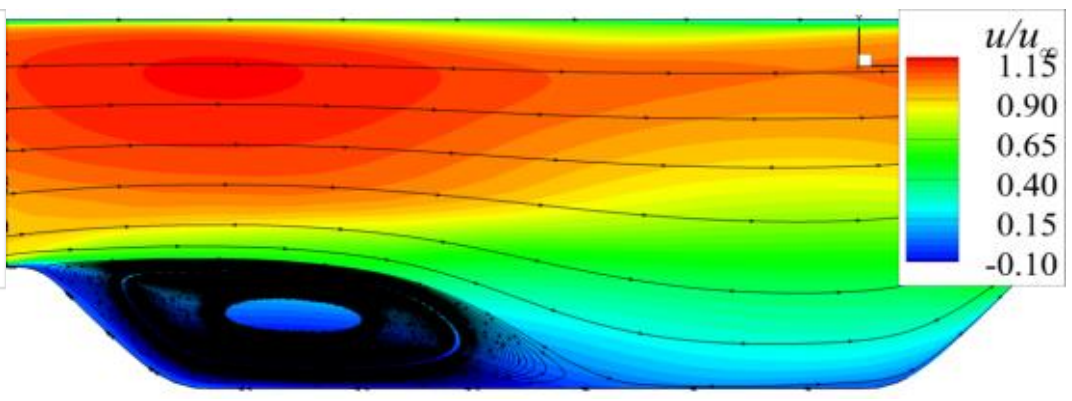

(2) FI, case 1

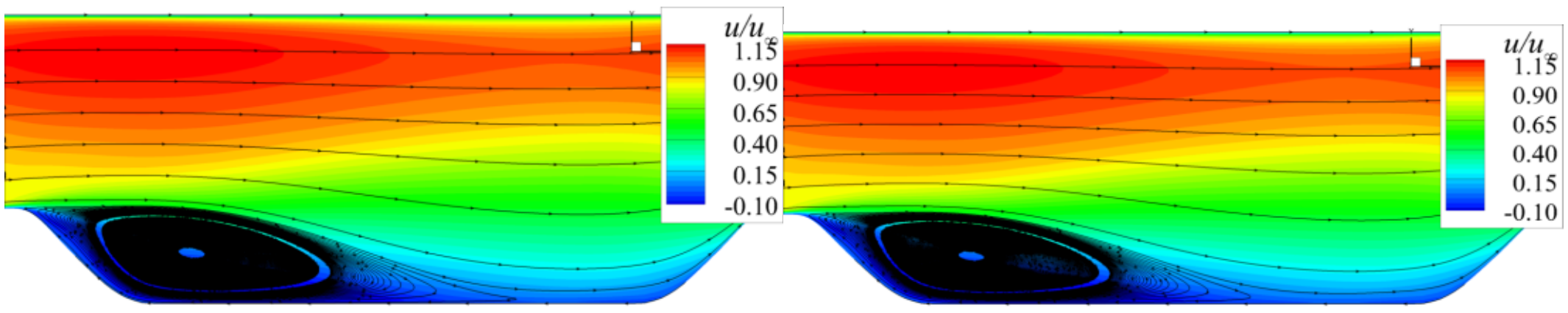

(4) FI, case 2 (4) FI, case 3

Fig. 17 Flow field from FI and DNS results

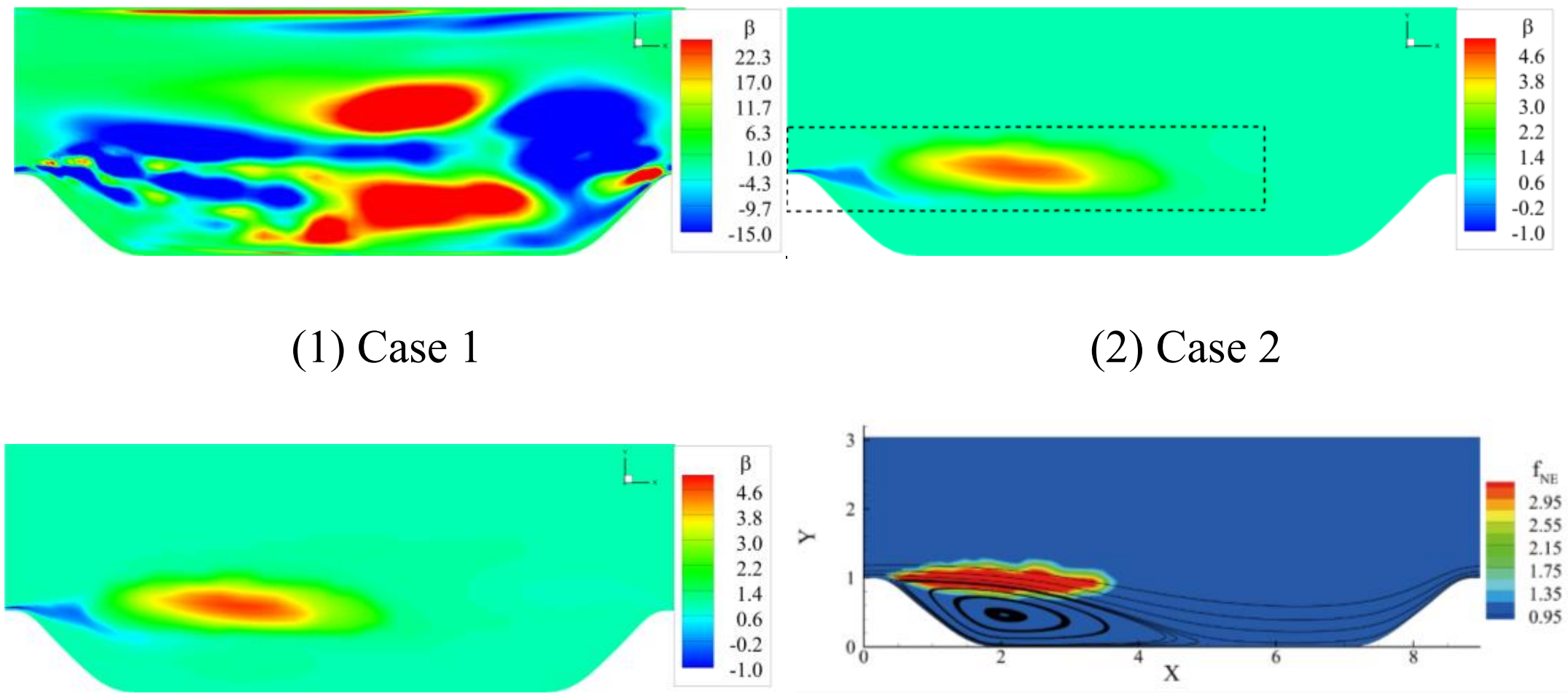

(1) Case 1 (2) Case 2

(3) Case 3 (4) Fixed region of the SPF $k-\overline{v^2}-\omega$ solution

Fig. 18 Correction fields from different cases

The correction field is also compared with the separation shear layer fixed region of the SPF $k-\overline{v^2}-\omega$ model [8]. The SPF $k-\overline{v^2}-\omega$ model makes use of the ratio of local turbulence production to dissipation to trigger an increase in nonequilibrium turbulence at the separation shear layer, and such modification can be represented by the term $f_{NE}$ [8]. Fig. 18(4) shows the resulting contour of $f_{NE}$ from the SPF $k-\overline{v^2}-\omega$ model. This shows that the modification of the nonequilibrium turbulence is triggered right in the separated shear region. For the FI, the corrected area is almost in the same region, as shown in Fig. 18(2) and Fig. 18(3), which further confirms the rationality of the physical mechanism of the flow FI.

### 3.2.2 Machine learning of the periodic hill case

As mentioned previously, there are several geometries of periodic hills with different steepness controlled by the parameter $\alpha$. Two FI cases with different geometries are used as the training dataset. Specifically, FIs are implemented in two cases with $\alpha$ being 0.5 and 0.8. There are 11200 samples (grid cells) in the dataset for training. Eighty percent of them are used as the training set, and the rest are used as the validation set. ANN is used as the ML model for turbulence model enhancement. The size of hidden layers is $(32, 16, 8)$. The hyperparameter settings for the ANN and the training process are the same as those for the S809 airfoil case.

The predicted correction fields using the features from the dataset are shown in Fig. 19. The locations and magnitude of the main corrections are shown to be predicted with high accuracy, which means that the performance of the ML model in offline prediction is satisfactory.

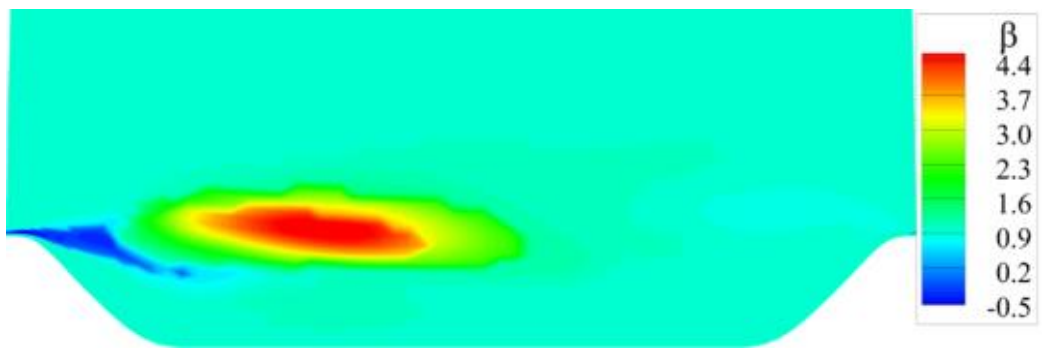

(1) $\alpha = 0.8$, label

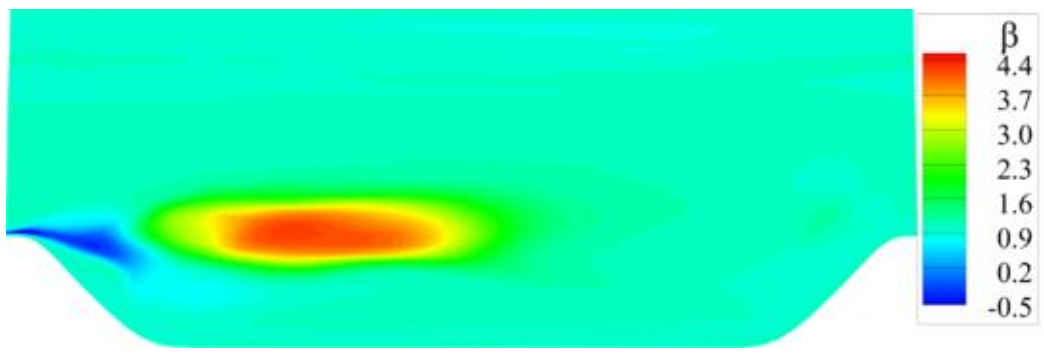

(2) $\alpha = 0.8$, prediction

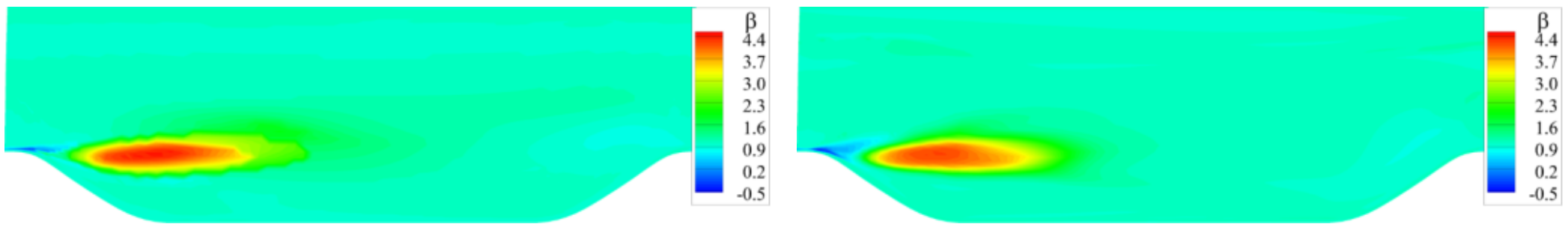

(1) α = 1.2, label (2) α = 1.2, prediction

Fig. 19 Predicted correction field using the features from the dataset

The trained ANN is embedded into the SA model to validate the performance of the machine-learning enhanced SA model. The enhanced model is first computed at the flow conditions the same as those of the FI. For the two cases of FI, the computed velocity profiles at different locations of the periodic hills are shown in Fig. 20. The velocity profiles from the enhanced SA model are close to the results from FI, and they are both closer to the DNS results than the original SA model. The computed flow fields are compared with the results from FI in Fig. 21. The ML augmented SA model can reproduce the result from FI with relatively high accuracy.

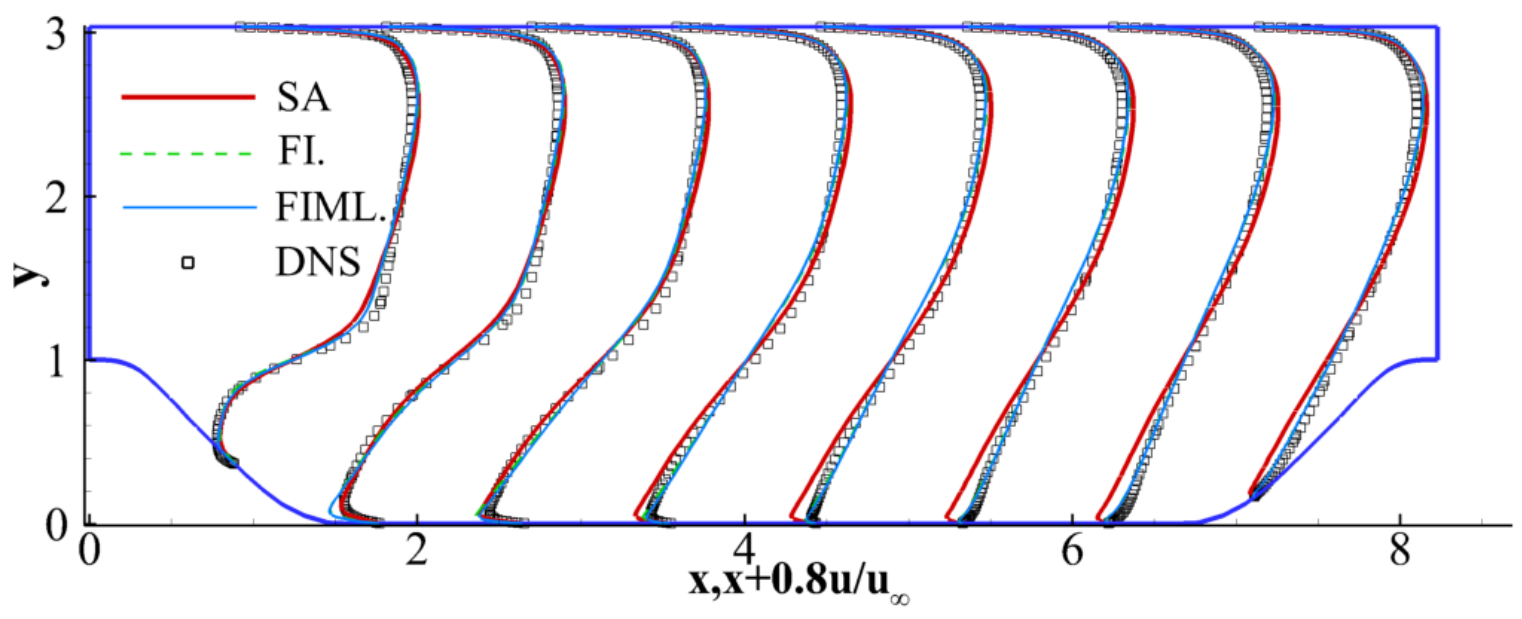

(1) α = 0.8

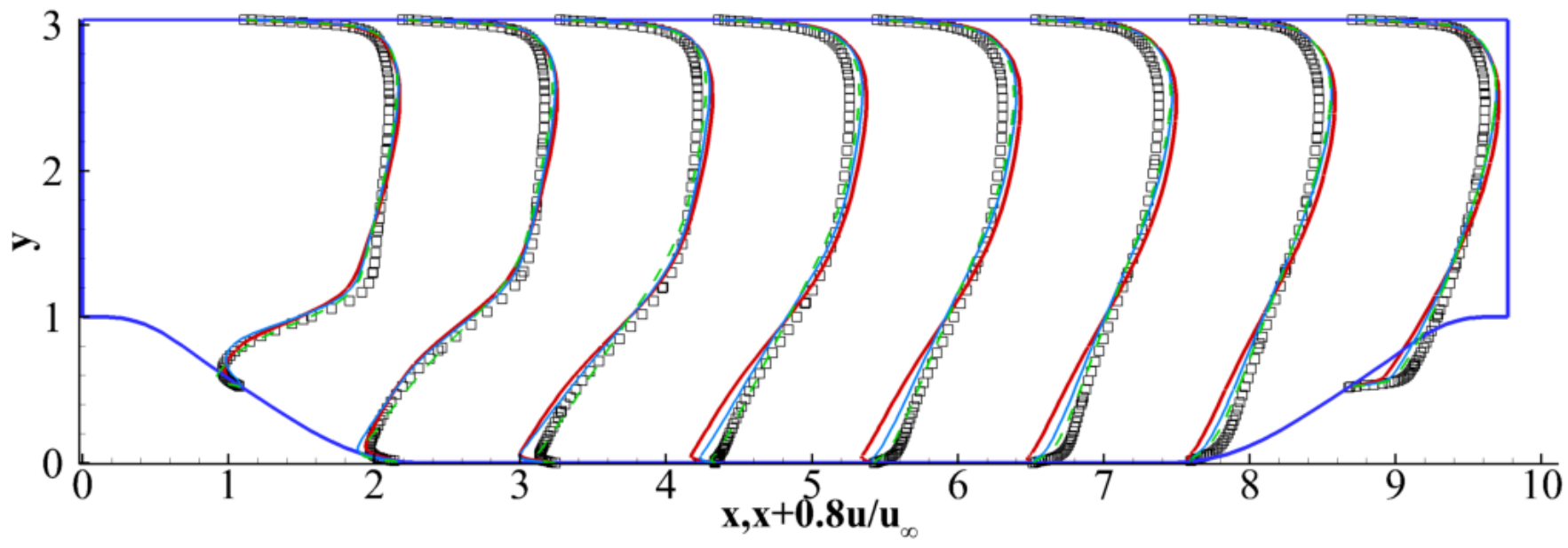

(1) $\alpha = 1.2$

Fig. 20 Computed velocity profiles at different locations

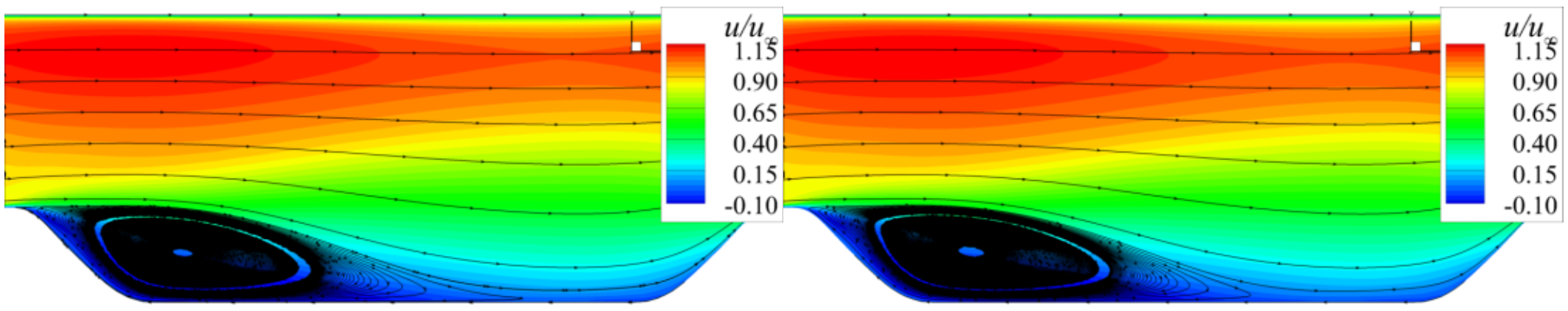

(1) $\alpha = 0.8$, ANN-augmented (2) $\alpha = 0.8$, FI

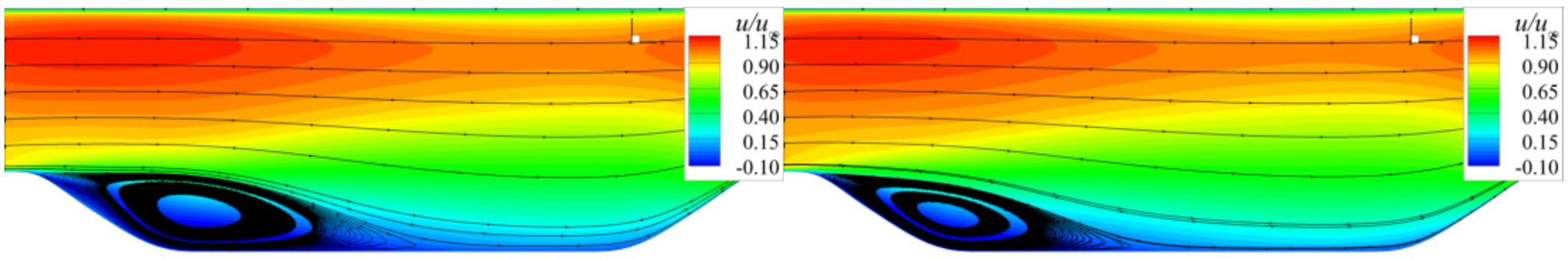

(1) $\alpha = 1.2$, ANN-augmented (2) $\alpha = 1.2$, FI

Fig. 21 Computed flow field compared with the results from FI

Finally, the enhanced SA model is tested at different periodic hill geometries to test the generalization ability. The case of $\alpha = 1.0$ is an interpolation case, and the cases of $\alpha = 0.5$ and $1.5$ are extrapolation cases. The computed velocity profiles at different locations are shown in Fig. 22. The results all show improvements compared with the original SA model, in which the velocity is corrected to be closer to the DNS results in most regions of the flow fields. For the cases of $\alpha = 0.5$ and 1.0, the velocity profiles are mainly modified near the lower wall downstream of the reattachment location, while for the case of $\alpha = 1.5$, the velocity profiles are mainly modified in the upper region of the flow. The ML-enhanced SA model extracts some generalized modeling knowledge from the training process, which helps improve the prediction of QoIs in unseen flow cases. However, the prediction of the whole flow field for the enhanced model can still be further promoted.

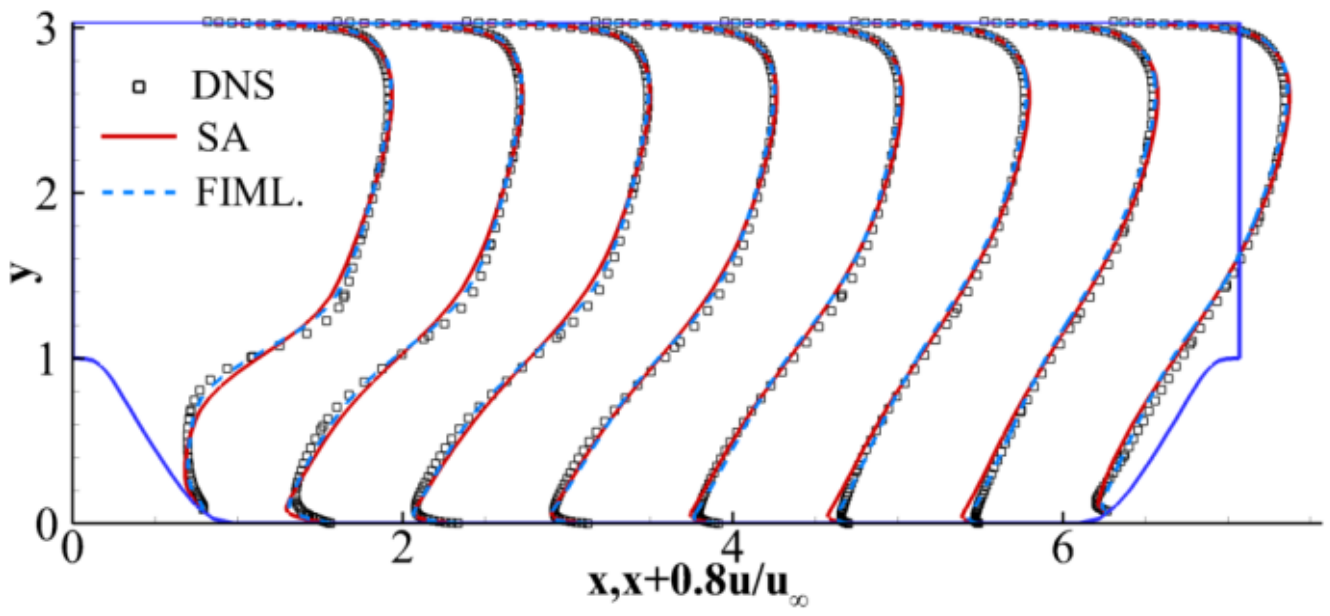

(1) $\alpha = 0.5$

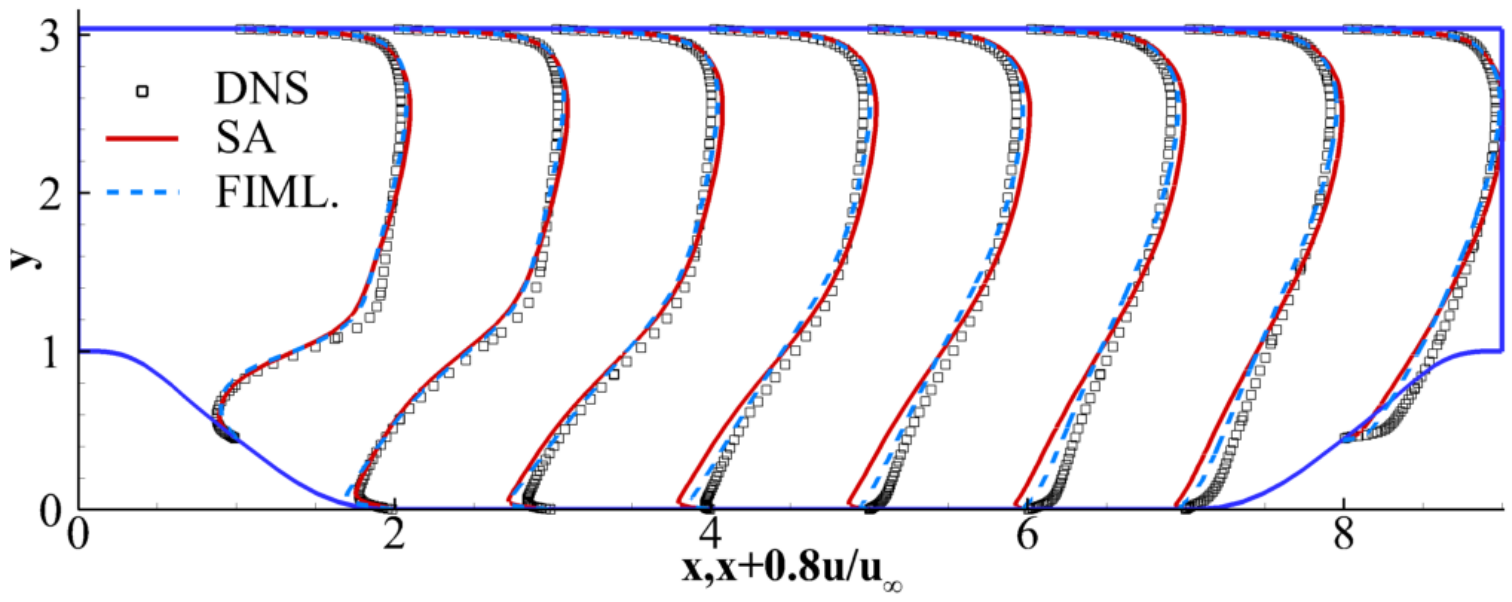

(2) $\alpha = 1.0$

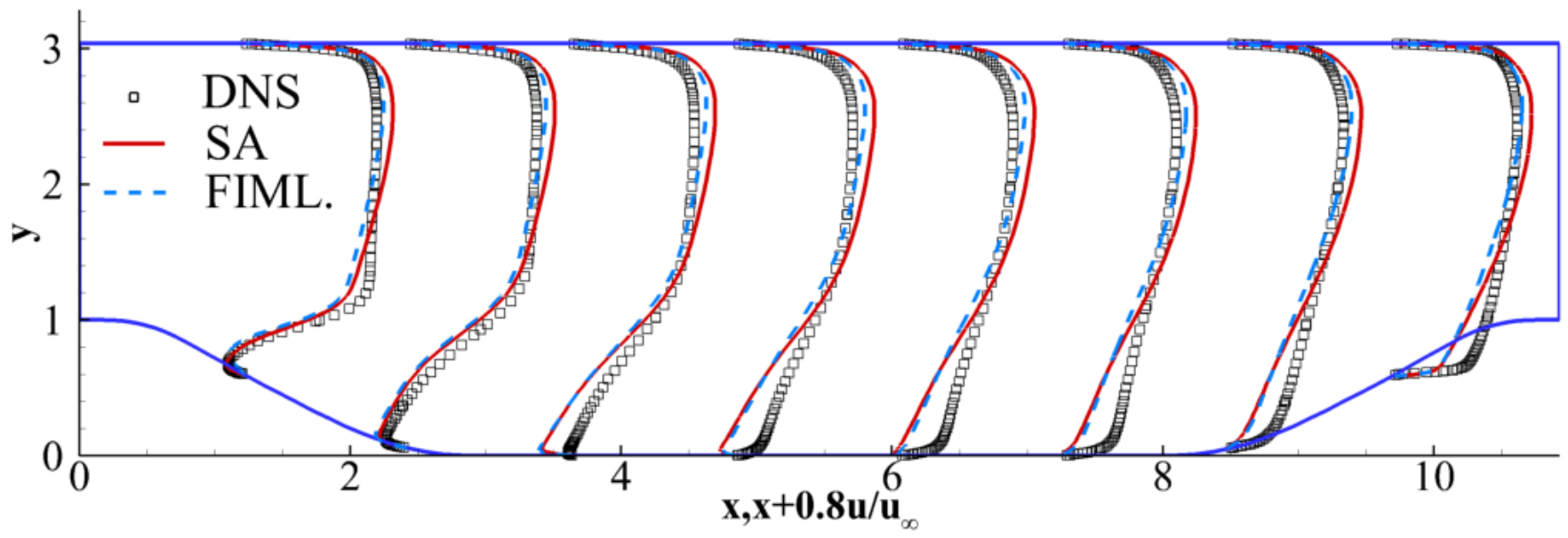

(3) $\alpha = 1.5$

Fig. 22 Computed velocity profiles at different locations

### *3.3 GLC305 airfoil with 944 ice shape*

Ice accretion on the wing causes severe degradation of aerodynamic performance, which threatens flight safety. Accurate simulation of the iced airfoil performance is important for aerodynamic design. Nevertheless, ice accretion on an airfoil makes it a nonstreamline body, which may induce flow separation. Current RANS models such as the SA model fail to make accurate predictions for iced airfoils near stall because of the lack of consideration for nonequilibrium effects in the separation shear flow region [8]. In this paper, FIML is implemented on airfoil GLC305 with ice shape 944 to modify the SA model. The experimental data used for FI were obtained from the NASA John H. Glenn Research Center at Lewis Field [49].

The geometry of the iced airfoil and the computation grid are shown in Fig. 23. There are 457 and 97 points along the streamwise and normal directions, respectively. The freestream condition is $Ma = 0.12, Re = 3.5 \times 10^6$. Previous studies have shown that the SA model tends to underestimate $C_L$ and $C_D$ at relatively large angles of attack [8], which needs to be corrected by the FIML approach in this paper.

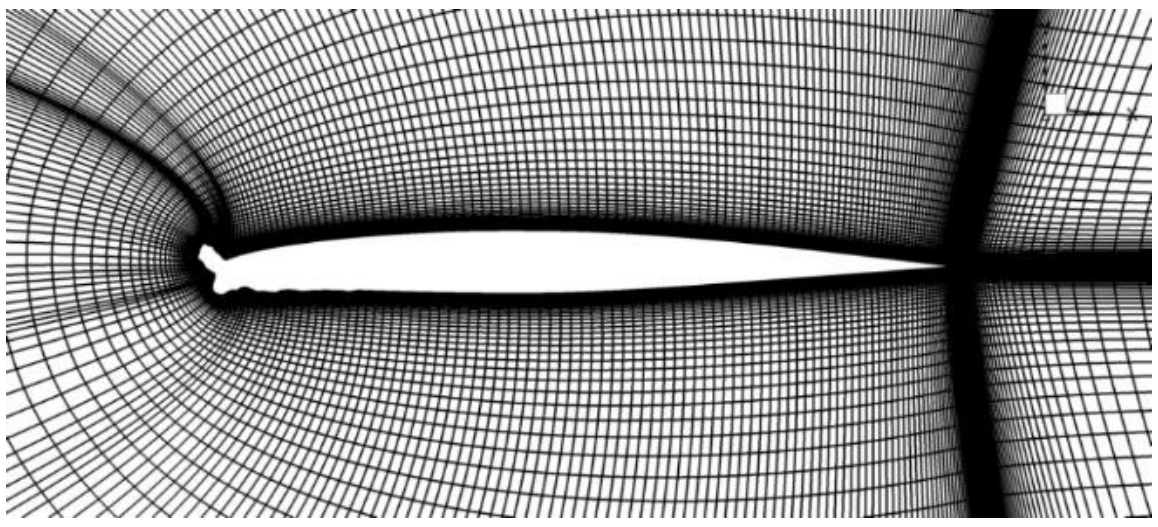

Fig. 23 Computation grid of the GLC305_944 airfoil

3.3.1 Field inversion

Two angles of attack, $4°$ and $6°$, are chosen for FI. Two tests of FI are implemented to study the effect of observation data on the FI results. The first case uses the lift coefficient $C_L$ and drag coefficient $C_D$ of the iced airfoil as the observation data, which is referred to as "FI1" in the following section. The objective function is written as:

$$J = \left[\left(C_L - h_L(\boldsymbol{\beta})\right)^2 + \left(C_D - h_D(\boldsymbol{\beta})\right)^2\right] + \sum_i \lambda \left(\beta_i - 1\right)^2 \quad (16)$$

The second case uses only the lift coefficient $C_L$ as the observation data, which is referred to as "FI2" in the following section. As shown in Eq. (16), the prior of $\boldsymbol{\beta}$ is Gaussian with uniform variance in space. The convergence histories of $C_L$ and $C_D$ during the FIs are shown in Fig. 24, and $C_L$ and $C_D$ computed by the SA model, FIs and experiment are compared in Table 3. The final results are highly consistent with the given target value.

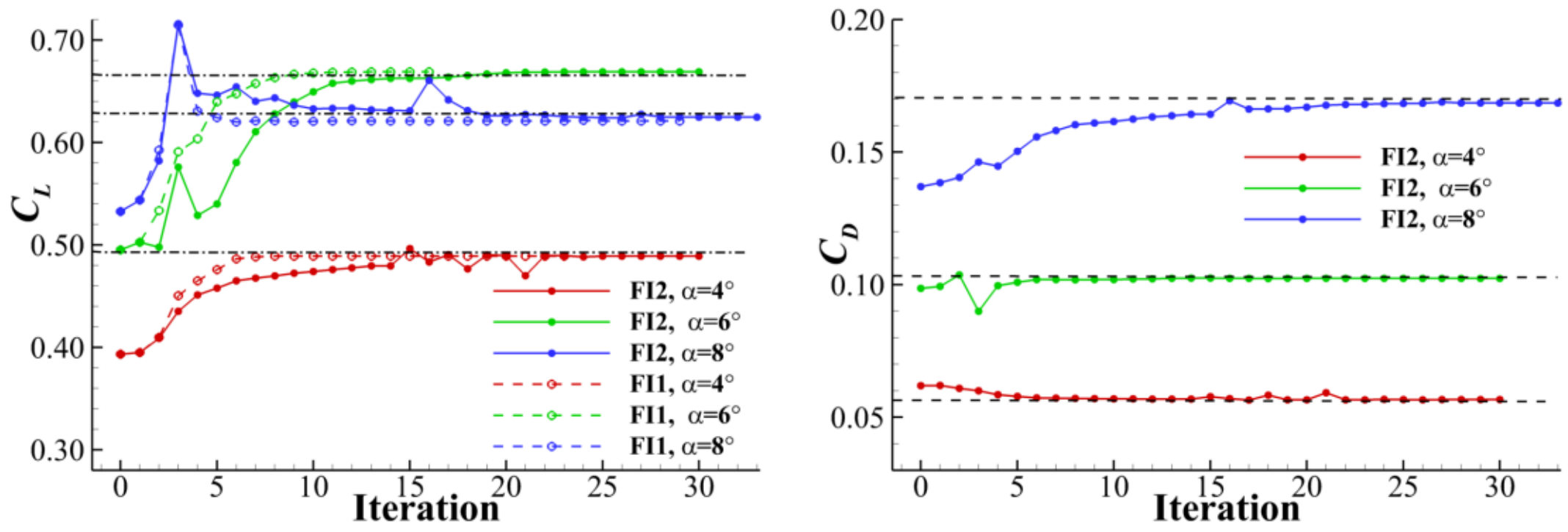

Fig. 24 Convergence of $C_L$ and $C_D$ during FIs

Table 3 $C_L$ and $C_D$ computed by the SA model, FIs and experiment

| Cases | 4° | | 6° | |
|---|---|---|---|---|
| | $C_L$ | $C_D$ | $C_L$ | $C_D$ |
| Original SA model | 0.393 | 0.0618 | 0.495 | 0.0986 |
| FI1 | 0.489 | 0.0566 | 0.669 | 0.1024 |
| FI2 | 0.489 | 0.0648 | 0.669 | 0.0950 |
| Experiment | 0.489 | 0.0566 | 0.669 | 0.1024 |

The flow fields before and after the FIs are shown in Fig. 25. The correction for the flow field is relatively small at an 4° angle of attack. However, at an angle of attack of 6°, the separation bubble predicted by the original SA model covers the whole airfoil, while the corrected bubbles take approximately half of the chord length. The flow reattachment after the shortened separation bubble helps to increase the lift coefficient $C_L$, which is closer to the experimental results in Table 3.

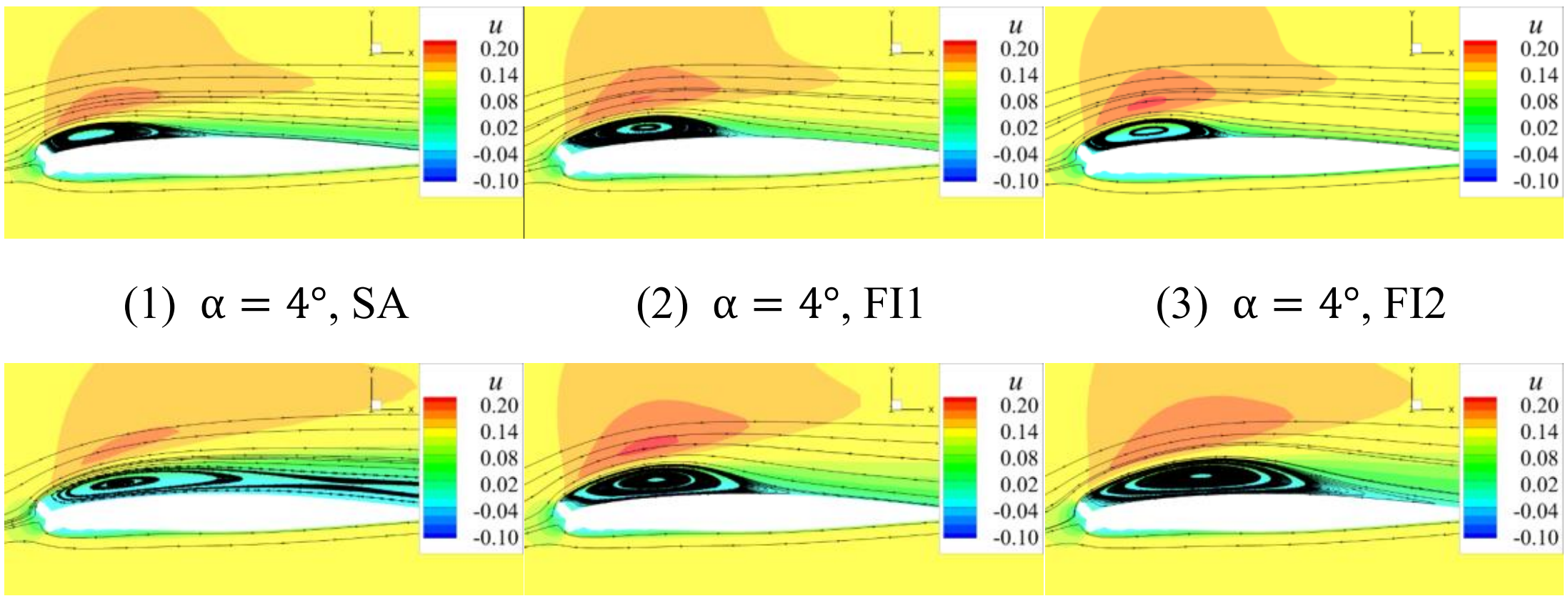

(1) α = 4°, SA (2) α = 4°, FI1 (3) α = 4°, FI2

(4) α = 6°, SA (5) α = 6°, FI1 (6) α = 6°, FI2

Fig. 25 Flow fields before and after the two FIs

The correction fields for the two FIs in different cases are shown in Fig. 26. The distributions of the correction fields show certain regularity. In front of the separation bubble, the multiplier of the production term is decreased, while in the back part of the separation bubble, the multiplier is increased. The results indicate that the main source of error of the original SA model comes from the separating shear flow region, and the turbulence mixing could be decreased or increased in the specific regions by the correction of FI. This is consistent with the analysis in previous studies.

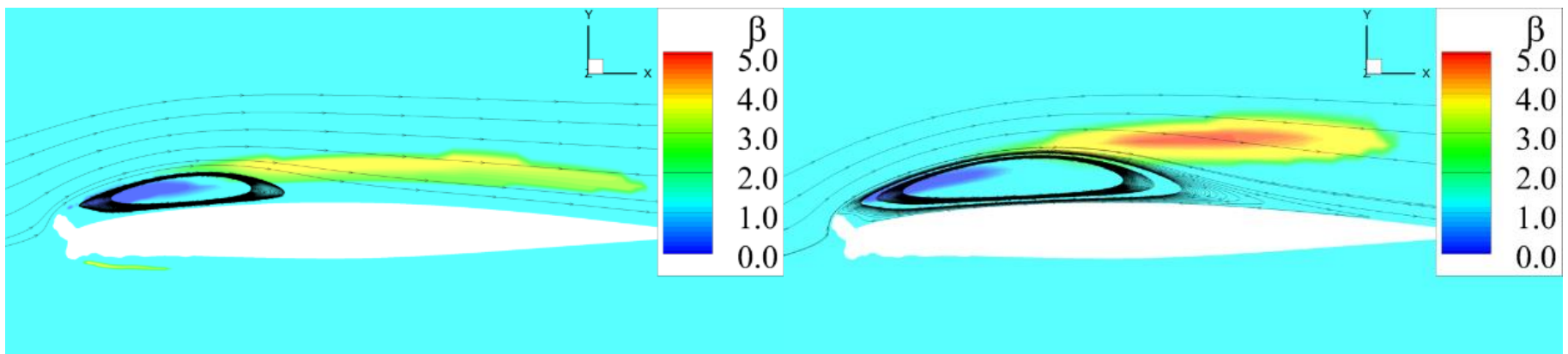

(1) α = 4°, FI1 (2) α = 6°, FI1

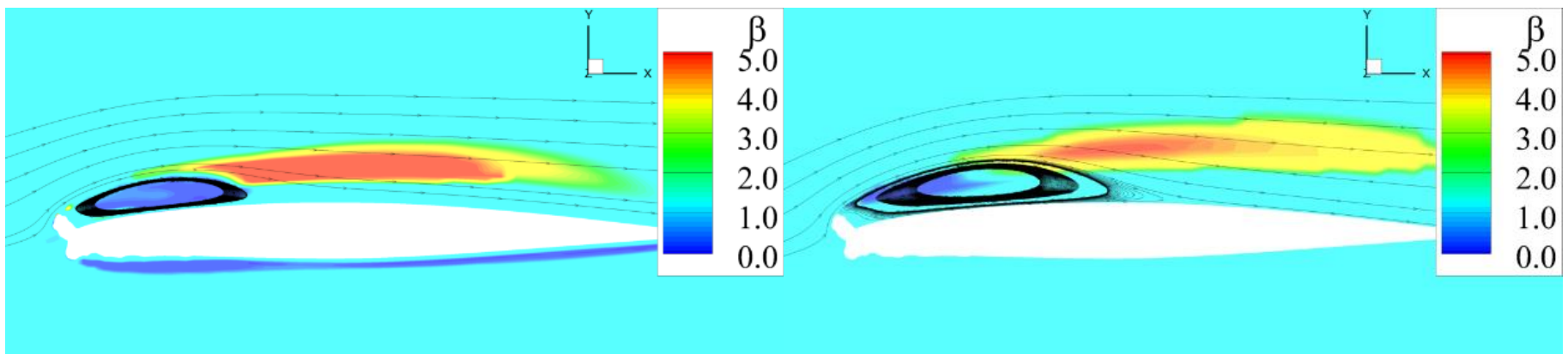

(4) $\alpha = 4°$, FI2 (5) $\alpha = 6°$, FI2

Fig. 26 Correction fields for the two FIs

The distribution of the correction is also compared with the SPF $\mathrm{k} - \overline{v^2} - \omega$ model [8], which made accurate predictions for flow around iced airfoils. As shown in Fig. 27, the SPF $\mathrm{k} - \overline{v^2} - \omega$ model activates the most important modification in the separating shear layer to account for the nonequilibrium characteristic, where the production-to-dissipation ratio of the turbulent kinetic energy ($P_k/\varepsilon$) is much greater than 1. The effect is similar to the increase in the production of eddy viscosity in the SA model. The nonequilibrium characteristic in separating shear layer is also found in periodic hill cases by both the SPF $\mathrm{k} - \overline{v^2} - \omega$ model and the FI in the last section. This demonstrates the physical realizability of the correction by the present FI. Moreover, both a decrease and an increase in turbulence mixing are obtained by FI, which shows the capability of more accurate inference for the specific flow case.

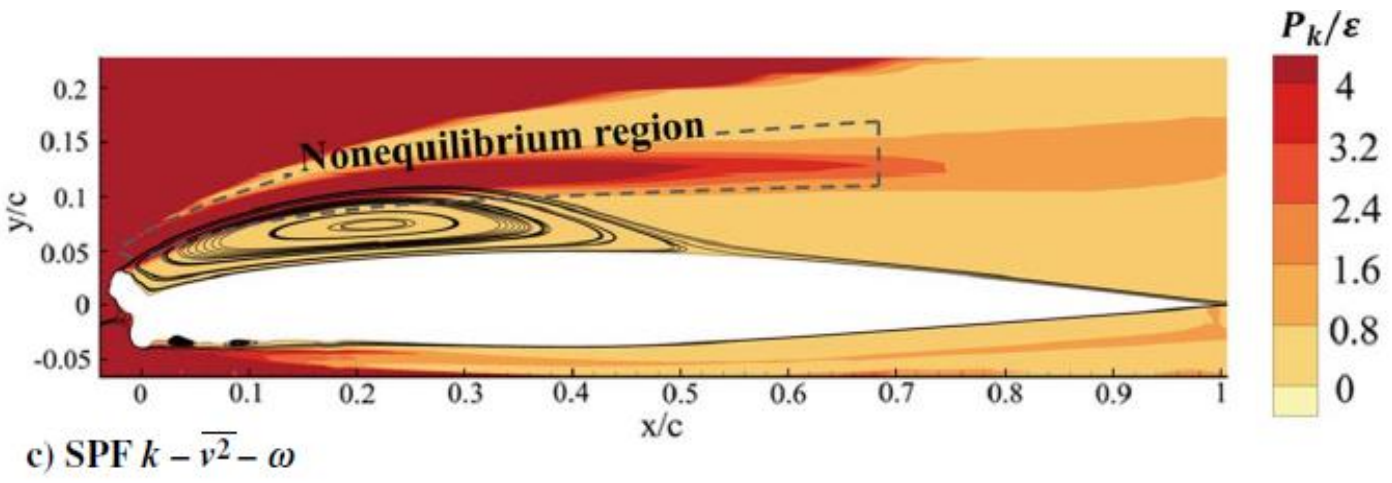

(1) $P_k/\varepsilon$

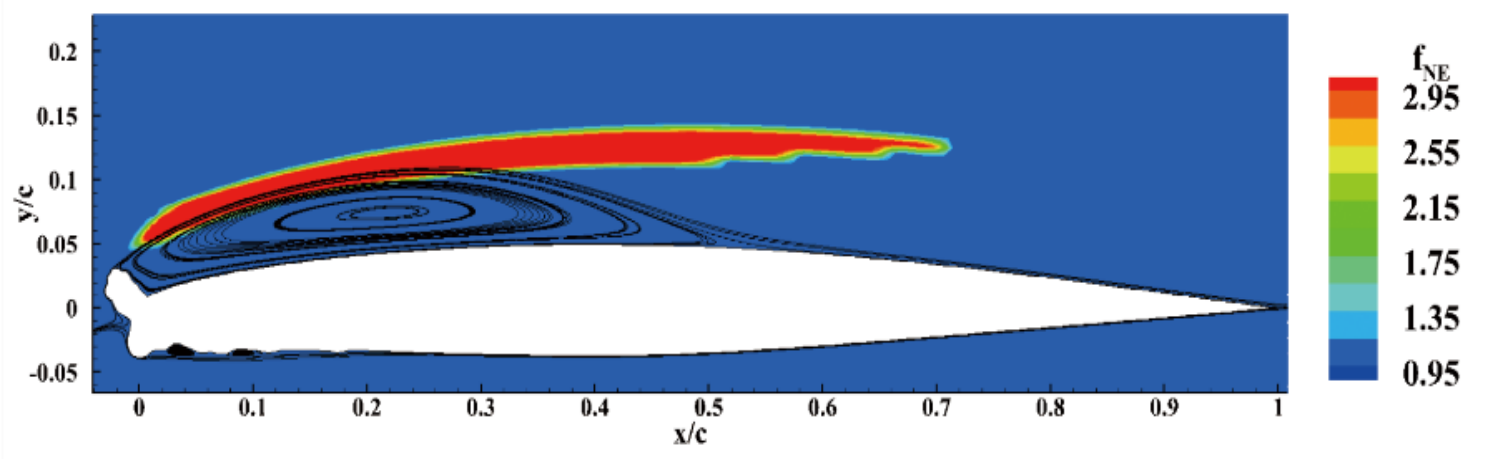

(2) $f_{NE}$

Fig. 27 Contours of $P_k/\varepsilon$ and the modifier $f_{NE}$ at the $6°$ angle of attack computed by the SPF $k-\overline{v^2}-\omega$ model [8]

The pressure coefficient ($C_p$) distributions on the airfoil by FI and the original SA model are shown in Fig. 28. There is usually a pressure platform in the separation region. Neither the height nor the length of the platform can be accurately predicted by the original SA model. Although the experimental $C_p$ distribution is not supplied to the inversion, the results of the two FIs are both closer to the experimental results than the original SA model. However, the $C_p$ distribution of $FI_1$ is better than the other. This is in line with our expectations: $FI_1$ has one more observation data point ($C_D$) than $FI_2$, so more information can be introduced to the inversion. More data from observations improve the credibility of data assimilation.

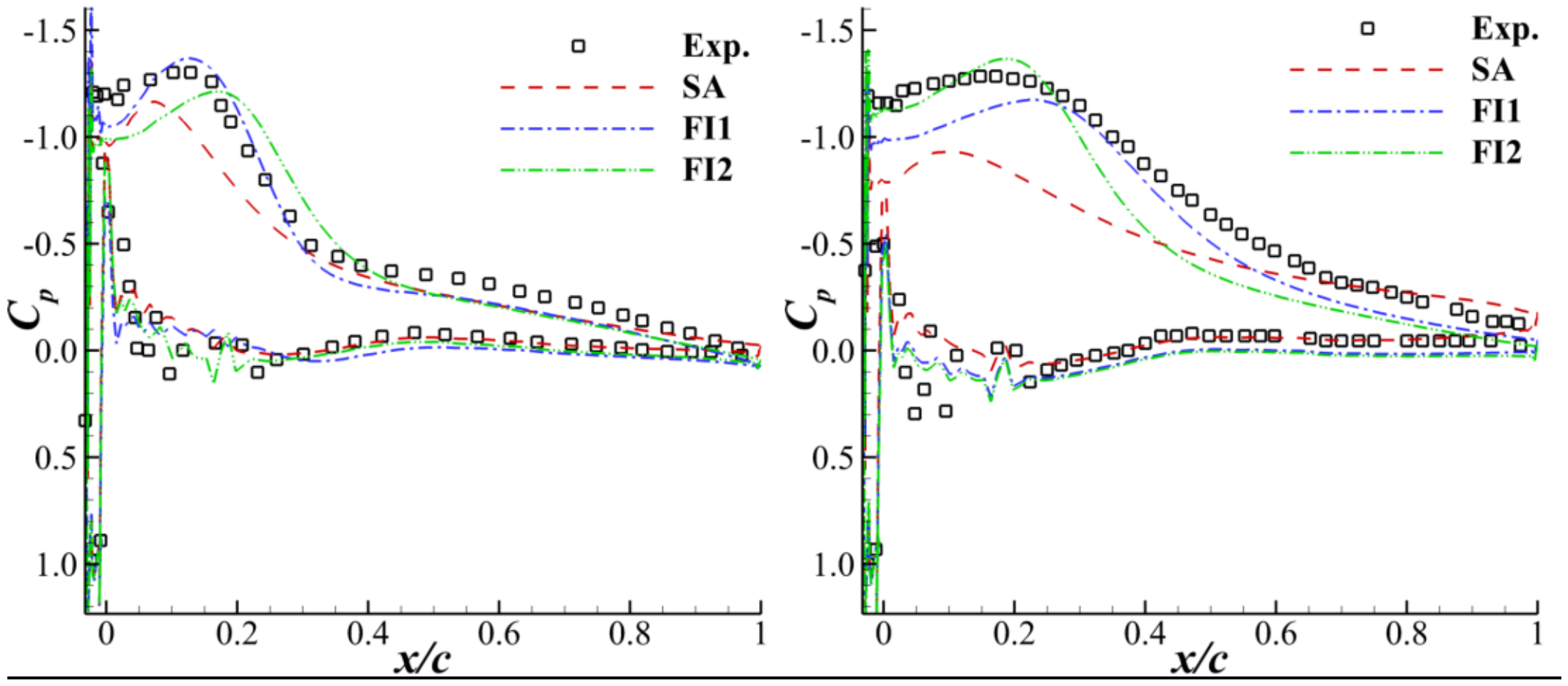

(1)4°　　　　(2)6°

Fig. 28 $C_p$ distribution on the airfoil by FI and the original SA model at different angles of attack

3.3.2 Machine learning for the iced airfoil case

The data from $\mathrm{FI}_1$ are used to train the ML model. According to experience, we chose an ANN with a hidden layer size of $(64, 32, 16)$ for training. Other configurations for the training are the same as those in section 3.1. The final flow field of FI at $4°$ and $6°$ angles of attack are combined to form the training dataset, in which there are a total of 131328 samples. Several local flow features at the $6°$ angle of attack are shown in Fig. 29.

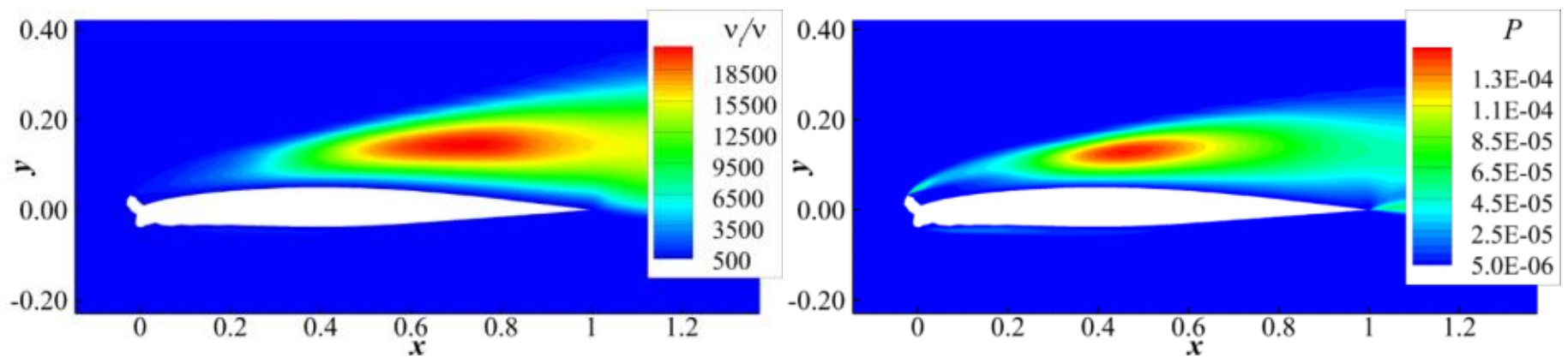

(1)$\frac{\tilde{\nu}}{\nu}$　　　　(2) $P$

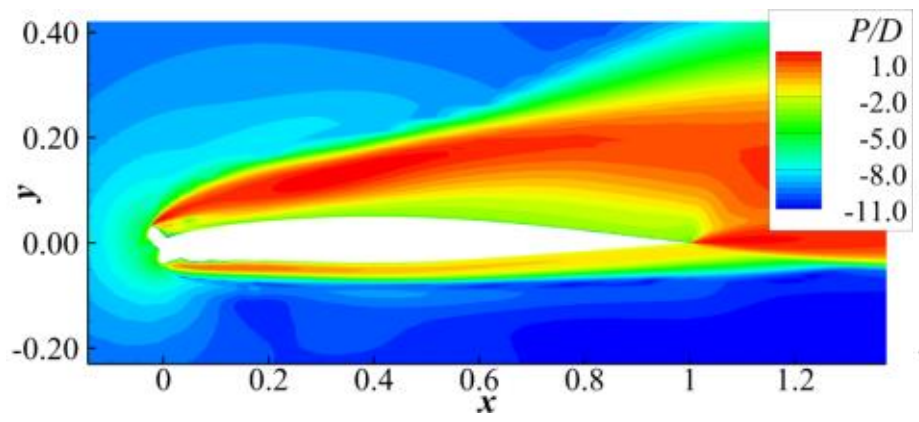

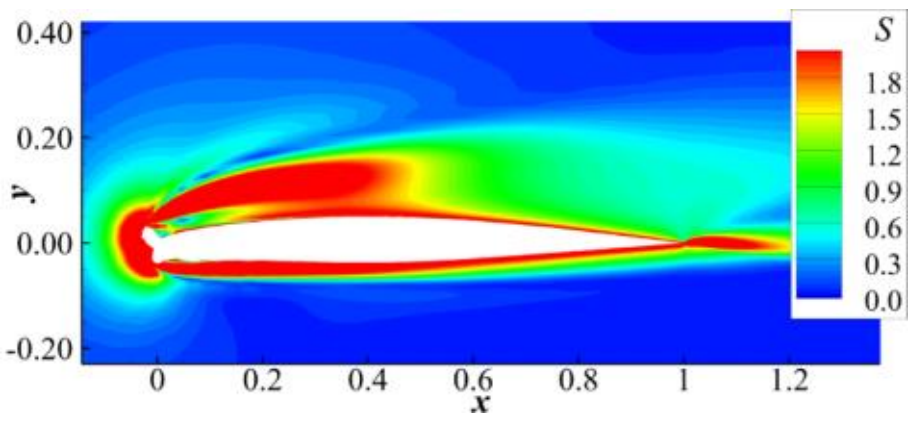

(3)$P/D$ (4) $S$

Fig. 29 Several flow features at a $6°$ angle of attack

The training is converged after 2000 epochs. As shown in Fig. 30, the MSE is sufficiently reduced during training. Next, the trained ANN is embedded into the original SA model to form the machine-learning enhanced SA model. The enhanced model is first validated at the same flow conditions as those for FI. The error of the computed $C_L$ and $C_D$ compared with those of experiments is shown in Table 4. The error of QoIs is significantly reduced by the enhanced SA model. The flow field from computation is shown in Fig. 31. Compared with the results from FI, the length and height of the separation bubbles are correctly reproduced. The computed $C_p$ distributions on the airfoil compared with FI and the original SA model are shown in Fig. 32. A comparison of Fig. 25 and Fig. 26 indicates that the results from FI can be reproduced with relatively high accuracy.

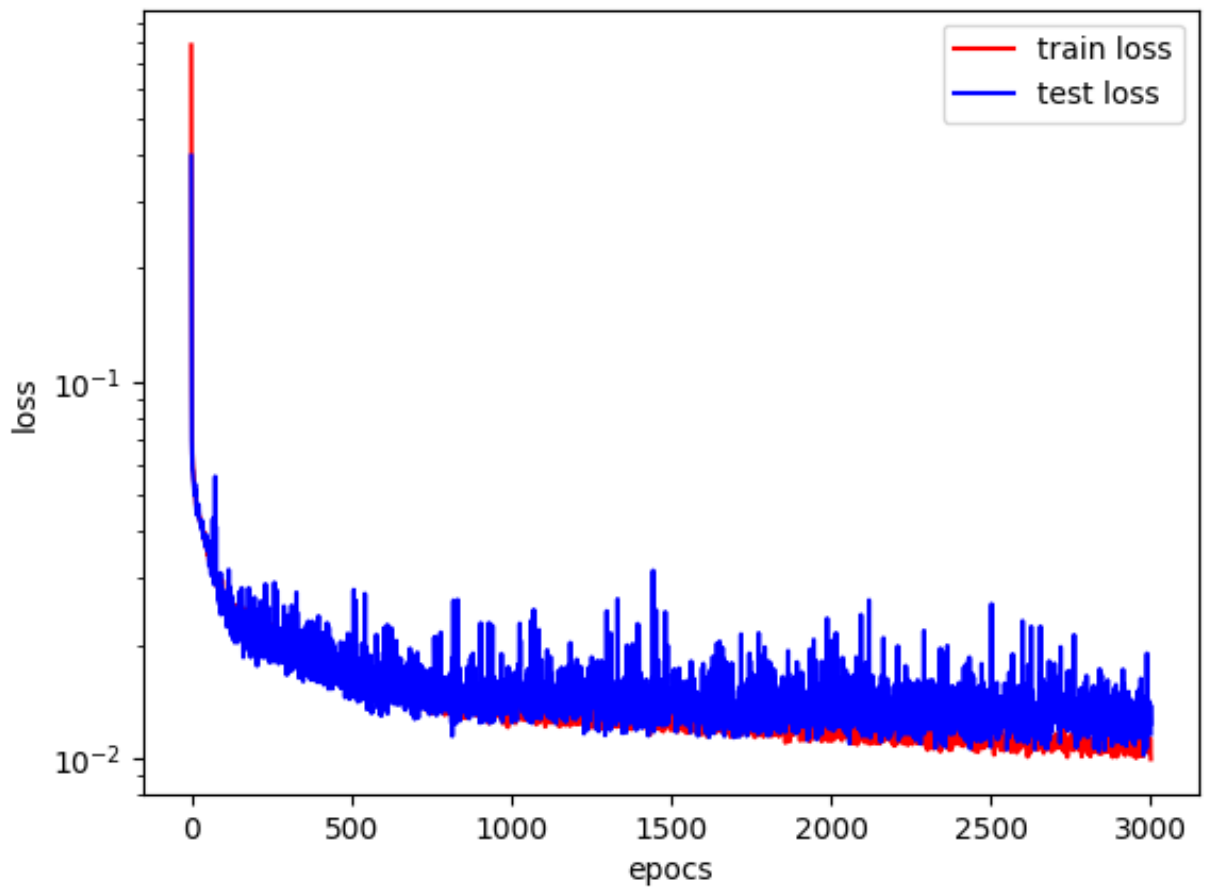

Fig. 30 Convergence of MSEs in the training and testing sets

Table 4 Error in the computed $C_L$ and $C_D$ with the original and augmented SA models

| α | Error in $C_L$ | | Error in $C_D$ | |
|---|---|---|---|---|
| | SA | Augmented SA | SA | Augmented SA |
| 4° | -21.5% | -7.0% | 6.4% | 3.5% |
| 6° | -26.8% | -5.5% | -7.1% | -6.3% |

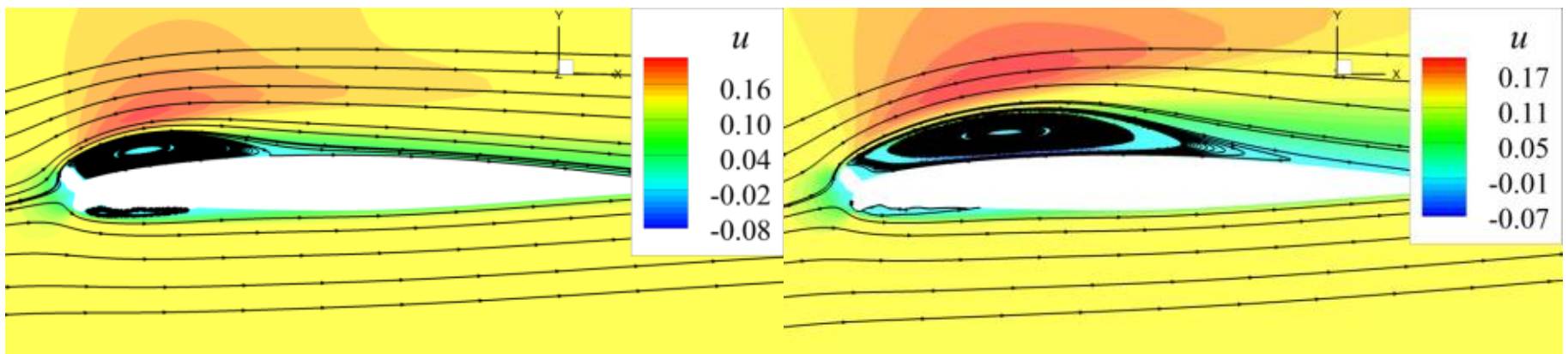

(1) $\alpha = 4°$ (2) $\alpha = 6°$

Fig. 31 Flow field computed by the enhanced SA model

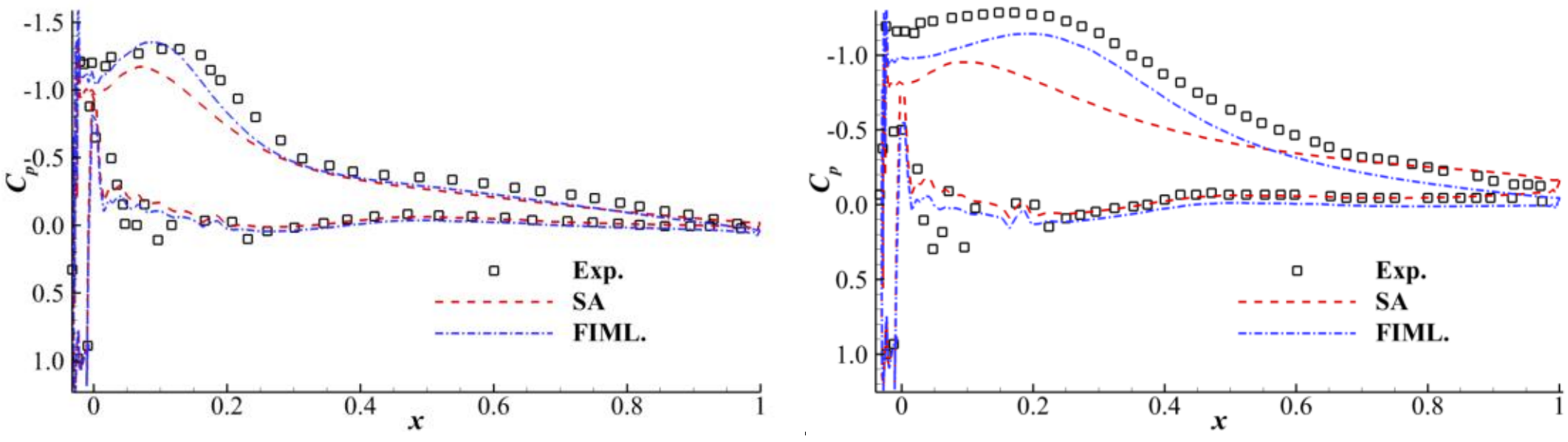

(1) $\alpha = 4°$ (2) $\alpha = 6°$

Fig. 32 Computed $C_p$ distributions on the airfoil compared with FI and the original SA model

Finally, the enhanced SA model is tested at a series of angles of attack from $0°$ to $7°$ to explore its generalization capability. The computed $C_L$ and $C_D$ are compared with the experimental results in Fig. 33. The enhanced SA model can predict $C_L$ with much higher accuracy than the original SA model at most angles of attack. In general, the enhanced SA model shows satisfying generalization capability in the linear range ($\alpha < 6°$) of the polar curve, while the results deteriorate at larger angles.

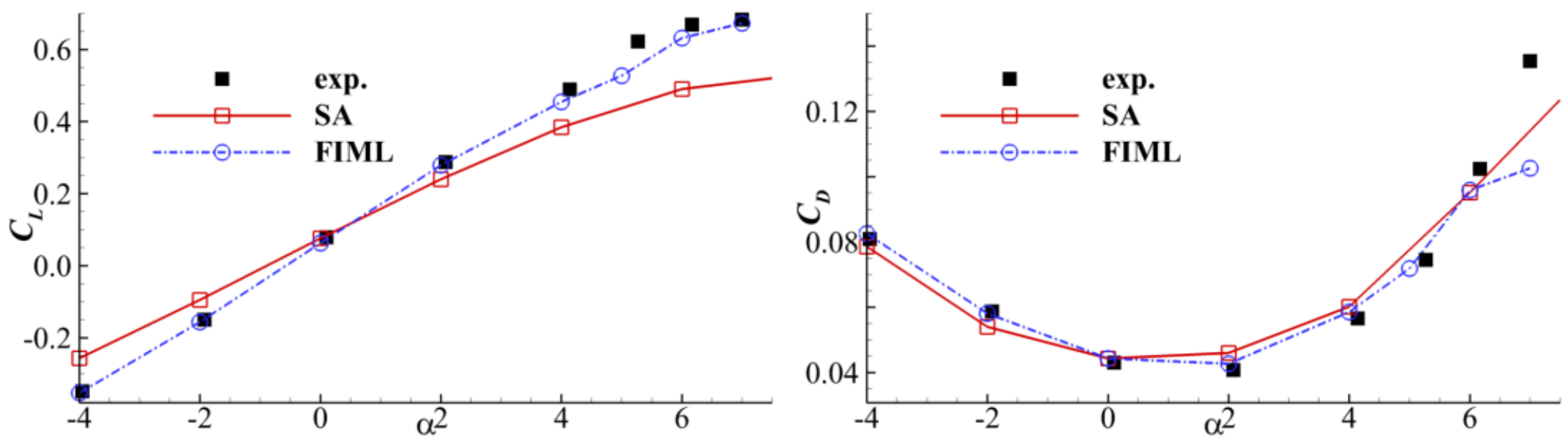

Fig. 33 Computed $C_L$ and $C_D$ by the enhanced and original SA models compared with the experimental results

## 4. Conclusions

Flow separation poses challenges to the precise modeling of turbulence. Previous studies have indicated that the model-form uncertainty in current RANS models mostly comes from the nonequilibrium effects related to the separation region. In this paper, the data-driven RANS modeling framework FIML is applied to the SA model to quantify and reconstruct the corrections related to nonequilibrium effects. The FIML framework is applied on three flow cases, namely, the S809 airfoil, the periodic hill, and the GLC305 airfoil with ice shape 944.

(1) FI is used to quantify the model-form uncertainty, with the prior of model uncertainty set manually using physical knowledge. The given observation data can be reproduced with significantly high accuracy after FI. In addition, the accuracy of the prediction of other QoIs can also be promoted even though they did not explicitly appear in the object function. Moreover, comparative analysis demonstrates that the correction made by the FI on iced airfoil and periodic hills accounts for the nonequilibrium effect in the separation shear layer, which is also a crucial consideration for the SPF $\mathrm{k}-\overline{v^2}-\omega$ model.

(2) The ML model is used to predict the model correction given local flow features. With the ANN trained with elaborately chosen hyperparameters, the machine-learning augmented SA model can reproduce the results from FIs with relatively high accuracy, and a certain extent of generalizability can be guaranteed at similar flow conditions. Nevertheless, the generalizability of the augmented turbulence model might be further promoted in future research.

## 5. Acknowledgments

This work was supported by the National Natural Science Foundation of China (grant nos. 91852108, 11872230, 92052203 and 91952302) and the Aeronautical Science Foundation of China (grant no. 2020Z006058002).

## 6. Conflict of interest statement

The authors declare that they have no conflicts of interest with regard to this work.

## References

[1] G B McCullough, D E Gault. Examples of three representative types of airfoil-section stall at low speed. NACA-TN-2502, 1951.

[2] C Marongiu, P L Vitagliano, G Zanazzi, and R Narducci. Aerodynamic Analysis of an Iced Airfoil at Medium/High Reynolds Number. AIAA Journal, 2008,

46(10):2469–2478. https://doi.org/10.2514/1.34550

[3] T B Gatski, C G Speziale. On explicit algebraic stress models for complex turbulent fows. Journal of fuid Mechanics, 1993, 254: 59-78.

[4] A Celic, E H Hirschel. Comparison of Eddy-Viscosity Turbulence Models in Flows with Adverse Pressure Gradient. AIAA Journal, 2006, 44(10):2156–2169. doi:10.2514/1.14902

[5] Tavoularis, S., Karnik, U., "Further experiments on the evolution of turbulent stresses and scales in uniformly sheared turbulence," Journal of Fluid Mechanics, Vol. 204, 1989, pp. 457-478. doi: 10.1017/S0022112089001837

[6] Rotta, J. C., "Turbulent Boundary Layers in Compressible Flow," Progress in Aerospace Sciences, Vol. 2, 1962, pp. 1-95. doi: 10.1016/0376-0421(62)90014-3

[7] L Fang, B Hkzc, B Lplc, et al. Quantitative description of non-equilibrium turbulent phenomena in compressors. Aerospace Science and Technology, 2017, 77:78-89.

[8] H R Li, Y F Zhang, H X Chen. Aerodynamic prediction of iced airfoils based on a modified three-equation turbulence model. AIAA Journal, 2020, 58(5): 3863-3876.

[9] H Li, Y Zhang, and H Chen. Numerical Simulation of Iced Wing Using Separating Shear Layer Fixed Turbulence Models, AIAA Journal, 2020.

[10] H Li, Y Zhang, and H Chen. Optimization design of airfoils under atmospheric icing conditions for UAV. Chinese Journal of Aeronautics, 2021. doi:

https://doi.org/10.1016/j.cja.2021.04.031

[11]C Rumsey. Exploring a Method for Improving Turbulent Separated-flow Predictions with k-ω Models. NASA/TM-2009-215952, 2009.

[12]K Duraisamy, G Iaccarino, H Xiao. Turbulence modeling in the age of data. Annual Review of Fluid Mechanics, 2019, 51: 357-377.

[13]M Emory, J Larsson, G Iaccarino. Modeling of structural uncertainties in Reynolds-averaged Navier-Stokes closures[J]. Physics of Fluids, 2013, 25(11):110822.

[14]J Ray, S Lefantzi, S Arunajatesan, et al. Learning an eddy viscosity model using Shrinkage and Bayesian calibration: A jet-in-crossflow case study. ASCE-ASME Journal of Risk and Uncertainty in Engineering Systems. Part B. Mechanical Engineering, 2018, 4(1): 011001

[15]H Kato, S Obayashi. Approach for uncertainty of turbulence modeling based on data assimilation technique. Computers & Fluids, 2013, 85: 2-7.

[16]H Xiao, J L Wu, J X Wang, et al. Quantifying and reducing model-form uncertainties in Reynolds-averaged Navier-Stokes simulations: A data-driven, physics-informed Bayesian approach. Journal of Computational Physics, 2016, 324: 115-136.

[17]T A Oliver, R D Moser. Bayesian uncertainty quantification applied to RANS turbulence models. Journal of Physics: Conference Series, 2011, 318(4): 042032.

[18] A P Singh, K Duraisamy. Using field inversion to quantify functional errors in turbulence closures. Physics of Fluids, 2016, 28: 045110.

[19] D P G Foures, N Dovetta, D Sipp, et al. A data-assimilation method for Reynolds-averaged Navier-Stokes-driven mean flow reconstruction. Journal of Fluid Mechanics, 2014, 759:404-431.

[20] Y Z Zhang, C Cheng, Y T Fan, et al. Data-driven correction of turbulence model with physics knowledge constrains in channel flow. Acta Aeronautica et Astronautica Ainica, 2020, 41(3): 123282.

[21] M Meldi, A Poux. A reduced order model based on Kalman filtering for sequential data assimilation of turbulent flows. Journal of Computational Physics ,2017, 347:207-234.

[22] C He, Y Liu, L Gan. Instantaneous pressure determination from unsteady velocity fields using adjoint-based sequential data assimilation. Physics of Fluids, 2020, 32(3): 035101.

[23] E J Parish, K Duraisamy. A paradigm for data-driven predictive modeling using field inversion and ma-chine learning. Journal of Computational Physics, 2015, 305: 758-774.

[24] B D Tracey, K Duraisamy, J J Alonso. A machine learning strategy to assist turbulence model development. Paper presented at AIAA Aerospace Sciences Meeting, 53rd, Kissimmee, FL, AIAA Pap. 2015-1287.

[25] J L Wu, H Xiao, E Paterson. Physics-Informed Machine Learning Approach for Augmenting Turbulence Models: A Comprehensive Framework. Physical Review Fluids, 2018, 3:074602.

[26] J LING, A KURZAWSKI, T J EMPLETON. Reynolds averaged turbulence modelling using deep neural networks with embedded invariance. Journal of Fluid Mechanics, 2016, 807: 155-166.

[27] K Duraisamy, Z J Zhang, A P Singh. New approaches in turbulence and transition modeling using datadriven techniques. Paper presented at AIAA Aerospace Sciences Meeting, 53rd, Kissimmee, FL, AIAA Pap. 2015-1284.

[28] A P Singh, S Medida, K Duraisamy. Machine-Learning-augmented predictive modeling of turbulent separated flows over airfoils. AIAA Journal, 2017, 55(7): 2215-2227.

[29] A Ferrero, A Iollo, F Larocca. Field inversion for data-augmented RANS modelling in turbomachinery flows. Computers & Fluids, 2020, 201:104474.

[30] P R Spalart, S R Allmaras. A one-equation turbulence model for aerodynamic flows. 30th Aerospace Sciences Meeting and Exhibit, 1992.

[31] J R Edwards, S Chandra. Comparison of Eddy Viscosity-Transport Turbulence Models for Three-Dimensional, Shock-Separated Flowfields. AIAA Journal, 1996, 34(4):756-763.

[32] P R Spalart. Strategies for Turbulence Modelling and Simulation. International

Journal of Heat and Fluid Flow, 2000, 21:252-263.

[33] O Pironneau. On optimum design in fluid mechanics. Journal of Fluid Mechanics, 1974, 64(1): 97-110.

[34] A Jameson. Aerodynamic design via control theory. Journal of Scientific Computing, 1988, 3(3): 377-401.

[35] J Elliott, J Peraire. Practical three-dimensional aerodynamic design and optimization using unstructured meshes. AIAA Journal, 1997, 35(9): 1479-1485.

[36] E Nielsen, W Anderson. Aerodynamic design optimization on unstructured meshes using the Navier-Stokes equations. AIAA Journal, 1999, 37(11): 957-964.

[37] G K W Kenway, C A Mader, P He, et al. Effective adjoint approaches for computational fluid dynamics. Progress in Aerospace Sciences, 2019, 110: 100542.

[38] L Hascoet, V Pascual. The Tapenade Automatic Differentiation tool: principles, model, and specification. ACM Transactions on Mathematical Software, Association for Computing Machinery, 2013, 39 (3), 10.1145/2450153.2450158. hal-00913983

[39] R Rashad, D W Zingg. Aerodynamic shape optimization for natural laminar flow using a discrete-adjoint approach. AIAA Journal, 2016, 54(11): 3321-3337.

[40] J X Wang, J L Wu, H Xiao. Physics-informed machine learning approach for reconstructing Reynolds stress modeling discrepancies based on DNS data. Physical Review Fluids, 2017, 2(3):1-22.

[41] M Schmelzer, R P Dwight, P Cinnella. Discovery of Algebraic Reynolds-Stress

Models Using Sparse Symbolic Regression. Flow Turbulence and Combustion, 2020, 104 (2-3): 579-603.

[42] Y Zhao, H D Akolekar, J Weatheritt, et al. turbulence model development using CFD-driven machine learning. Journal of Computational Physics, 2020, 411:109413.

[43] D M Somers. Design and Experimental Results for the S809 Airfoil. Renewable Energy Lab. Rept. NREL/SR 440-6918, Golden, CO, 1997.

[44] J R Holland, J D Baeder, K Duraisamy. Field Inversion and Machine Learning With Embedded Neural Networks: Physics-Consistent Neural Network Training. AIAA Aviation 2019 Forum, 17-21 June 2019, Dallas, Texas.

[45] A Paszke, S Gross, F Massa, et al. PyTorch: An Imperative Style, High-Performance Deep Learning Library. 33rd Conference on Neural Information Processing Systems (NeurIPS 2019), Vancouver, Canada.

[46] H Xiao, P Cinnella. Quantification of model uncertainty in RANS simulations: A review. Progress in Aerospace Sciences, 2019, 108: 1-31.

[47] Y Yin, P Yang, Y Zhang, et al. Feature selection and processing of turbulence modeling based on an artificial neural network. Physics of Fluids, 2020, 32(10):105117.

[48] H Xiao, J Wu, S Laizet, and L Duan. Flows over periodic hills of parameterized geometries: A dataset for data-driven turbulence modeling from direct simulations. Computers & Fluids, 2020, 200, 104431.

[49] A P Broeren, M B Bragg, H E Addy. Flowfield measurements about an airfoil

with leading-edge ice shapes. Journal of Aircraft, 2006, 43(4): 1226-1234.